\documentclass[reprint,
superscriptaddress,
showpacs,
nofootinbib,
prd,
aps,
floatfix,
]{revtex4-1}

\usepackage{amsmath,amssymb,amsfonts,amsthm}
\usepackage[english]{babel}
\usepackage{url}
\usepackage{bm}
\usepackage[latin1]{inputenc}
\usepackage{graphicx,rotating}
\usepackage[colorlinks=true,linkcolor=black, citecolor=blue]{hyperref}

\usepackage{mathtools}
\newcommand{\slashed}[1]{\ensuremath{\mathrlap{\!\not{\phantom{#1}}}#1}}

\begin{document}
\title{Standard Model \texorpdfstring{$\mathcal{O}(\alpha)$}{O(alpha)} renormalization of \texorpdfstring{$g_A$}{gA} and its impact on new physics searches}
\author{Leendert Hayen}
\email[]{lmhayen@ncsu.edu}
\affiliation{Department of Physics, North Carolina State University, Raleigh, North Carolina 27695, USA}
\affiliation{Triangle Universities Nuclear Laboratory, Durham, North Carolina 27708, USA}
\affiliation{Instituut voor Kern- en Stralingsfysica, KU Leuven, Celestijnenlaan 200D, B-3001 Leuven, Belgium}

\date{\today}
\begin{abstract}
We present an $\mathcal{O}(\alpha)$ Standard Model calculation of the inner radiative corrections to Gamow-Teller $\beta$ decays. We find that \textit{a priori} contributions arise from the photonic vertex correction and $\gamma W$ box diagram. Upon evaluation most elastic contributions vanish due to crossing symmetry or cancellation between isoscalar and isovector photonic contributions, leaving only the polarized parity-odd contribution, i.e., the Gamow-Teller equivalent of the well-known axial $\gamma W$ box contribution for Fermi decays. We show that weak magnetism contributes significantly to the Born amplitude, and consider additional hadronic contributions at low energy using a holomorphic continuation of the polarized Bjorken sum rule constrained by experimental data. We perform the same procedure for the Fermi inner radiative correction through a combination of the running of Bjorken and Gross-Llewellyn Smith sum rules. We discuss heavy flavor, higher-twist, and target mass corrections and find a significant increase at low momentum from the latter. We find $\Delta_R^A = 0.02532(22)$ and $\Delta_R^V = 0.02473(27)$ for axial and vector inner radiative corrections, respectively, resulting in $\Delta_R^A-\Delta_R^V=0.60(5) \times 10^{-3}$, which allows us to extract $g_A^0$ for the first time to our knowledge. We discuss consequences for comparing experimental data to lattice calculations in beyond Standard Model fits. Further, we show how some traditional $\beta$ decay calculations contain part of this effect but fail to account for cancellations in the full $\mathcal{O}(\alpha)$result. Finally, we correct for a double-counting instance in the isospin $T=1/2$ mirror decay extraction of $|V_{ud}|$, the up-down matrix element of the Cabibbo-Kobayashi-Maskawa matrix, resolving a long-standing tension and leading to increased precision.
We discuss consequences for comparing experimental data to lattice calculations in Beyond Standard Model fits. Further, we show how some traditional $\beta$ decay calculations contain part of this effect but fail to account for cancellations in the full $\mathcal{O}(\alpha)$ result. Finally, we correct for a double-counting instance in the isospin $T=1/2$ mirror decay extraction of $|V_{ud}|$, the up-down matrix element of the Cabibo-Kobayashi-Maskawa matrix element, resolving a long-standing tension and leading to increased precision.
\end{abstract}

\maketitle


\section{Introduction}
\label{sec:introduction}
Precision studies of neutron and nuclear $\beta$ decays were of paramount importance in the construction of the Standard Model and provide stringent constraints on TeV-scale Beyond Standard Model (BSM) physics \cite{Renton1990, Commins1983, Holstein2014, Cirigliano2013b, Gonzalez-Alonso2018}. Electroweak radiative corrections (EWRC) play a central role in this endeavor \cite{Czarnecki2004, Sirlin2013}, and require to be known to high precision. This is particularly so for top-row unitarity tests of the Cabibo-Maskawa-Kobayashi (CKM) matrix \cite{Abele2002, Towner2010a, Czarnecki2018, Czarnecki2020}, where the final uncertainty is dominated by that on EWRC for some systems. Recently, new theoretical work on radiative corrections common to neutron and superallowed Fermi decays \cite{Seng2018, Seng2019b, Gorchtein2018, Seng2020} has caused a reevaluation of older work \cite{Marciano2006, Czarnecki2019} and an apparent discrepancy with CKM top-row unitarity.

Following several new experimental results \cite{Pattie2018, Brown2018, Markisch2019, Beck2019}, the neutron is quickly reaching competitive levels with superallowed $\beta$ decays \cite{Hardy2015, Hardy2020} for an extraction of $|V_{ud}|$, the up-down CKM matrix element through
\begin{equation}
    |V_{ud}|^2\tau_n\left(f_V+3f_A\lambda^2\right) = \frac{2\pi^3}{G_F^2m_e^5g_V^2}\frac{1}{1+RC}
    \label{eq:general_tiple_relation}
\end{equation}
where $\tau_n$ is the neutron lifetime, $G_F \approx 10^{-5}$ GeV$^{-2}$ is the Fermi coupling constant, $m_e$ is the electron mass, $\lambda \equiv g_A/g_V$ is the ratio of axial and vector coupling constants, $f_{V/A}$ their respective phase space integrals, and $RC$ represents electroweak radiative corrections \cite{Czarnecki2018}. The latter is traditionally written as
\begin{equation}
    1 + RC = 1 + \delta_\text{out}(E) + \Delta_R^V
    \label{eq:RC_def}
\end{equation}
where $\delta_\mathrm{out}(E)$ is an energy dependent, but nuclear structure independent correction and $\Delta_R^V$ is the so-called inner radiative correction for the vector charged current, i.e., a renormalization of $g_V$ \cite{Czarnecki2019, Seng2018, Seng2019b}. While the latter is protected from QCD corrections through the Ademollo-Gatto theorem \cite{Ademollo1964}, the axial-vector coupling constant, $g_A$, receives both strong and electroweak corrections at next-to-leading order. As these bring significant complexity, however, one typically continues with an experimentally obtained value that contains all further corrections. In other words, $g_A$ from Eq. (\ref{eq:general_tiple_relation}) is commonly defined as
\begin{equation}
    g_A^\mathrm{eff} = g_A^\text{QCD}\left[1 + \frac{1}{2}\left(\Delta_R^A - \Delta_R^V\right) + \delta_\text{BSM} \right]
    \label{eq:gA_def_QCD}
\end{equation}
where $g_A^\mathrm{QCD}$ contains strong interaction effects, $\Delta_R^A$ are electroweak corrections to $g_A$, and we have explicitly allowed the possibility for BSM interference.

Following great progress from lattice QCD (LQCD) in the past years \cite{Chang2018, Gupta2018, Aoki2020}, a comparison between an experimental $g_A^\mathrm{eff}$ and theoretical $g_A^\text{QCD}$ results has become a new, clean channel for probing right-handed currents in the electroweak sector \cite{Alioli2017, Gonzalez-Alonso2018}. Specifically, if one assumes that the bulk of the electroweak corrections are common to both $g_V$ and $g_A$, $\Delta_R^A-\Delta_R^V$ is small and ${g_A^\mathrm{eff} \approx g_A^\mathrm{QCD}(1+\delta_\mathrm{BSM})}$. Up to now, the difference in vector and axial-vector EWRC has been assumed to be smaller than $0.1\%$, although no complete calculations have been performed \cite{Sirlin1968, Garcia1983, Kurylov2002, Kurylov2003, Fukugita2004}.

Here, we focus on a Standard Model $\mathcal{O}(\alpha)$ calculation of $\Delta_R^A$. The paper is organized as follows. Section \ref{sec:outline} provides a sketch of what physics enters the calculation of $RC$ in Eq. (\ref{eq:RC_def}), and discusses the tools we will be using. In Sec. \ref{sec:vertex_correction}, we treat the Standard Model electroweak vertex correction, followed by Sec. \ref{sec:gamma_W_box} where we discuss the box diagrams. These findings coalesce into Sec. \ref{sec:summary_order_alpha} which summarizes the effective nucleon couplings and nuclear effects. Finally, we discuss two consequences of our findings in Secs. \ref{sec:gA_BSM} and \ref{sec:traditional_beta_decay}, treating the comparison to LQCD and consistency errors in traditional $\beta$ decay formalisms and mirror $|V_{ud}|$ extraction, respectively.

\section{Overview of Standard Model input}
\label{sec:outline}
Before we proceed, we sketch some general outlines of the problem. For a more general discussion, we refer the reader to several excellent reviews \cite{Sirlin1978, Sirlin2013, Wilkinson1982, Wilkinson1995b, Wilkinson1997, Wilkinson1998}. 

\subsection{Sketch of the ingredients}
The $\mathcal{O}(\alpha)$ radiative corrections (RC) to the Standard Model $\beta$ decay amplitude at first sight correspond to a large number of contributing diagrams, ranging from virtual electroweak boson exchange to Higgs interactions \cite{Sirlin1978}. Many of these, however, contribute only to $\mathcal{O}(G_F^2)$ upon evaluation, and the final selection is much more modest. Here we are interested only in those which can differ between Fermi and Gamow-Teller transitions, so that all diagrams which leave the interaction vertex unaltered (wave function renormalization, $\mathcal{O}(\alpha)$ bremsstrahlung, etc.) serve only to guarantee gauge invariance in the evaluation of Eq. (\ref{eq:gA_def_QCD}) and remove IR divergences.

We start with the description of the theoretically clean muon $\beta$ decay, which was one of the early successes for the calculation of EWRC \cite{Kinoshita1959}. Specifically, one found that using the older $V$-$A$ current-current interaction,
\begin{equation}
\mathcal{H}_\beta = \frac{G_F}{\sqrt{2}}\bar{e}\gamma^\lambda(1-\gamma^5)\mu \times \bar{\nu}_\mu\gamma_\lambda(1-\gamma^5)\nu_e + \mathrm{h.c.} 
\label{eq:fermi_cc_interaction}
\end{equation}
with $G_F$ the so-called Fermi coupling constant, the radiative corrections were both infrared (IR) and ultraviolet (UV) finite. In this theory the only gauge boson that is present is the photon, and the muon lifetime could be cleanly calculated to $\mathcal{O}(\alpha)$ with $\alpha$ the fine-structure constant
\begin{equation}
    \frac{1}{\tau_\mu} = \frac{G_F^2m_\mu^5}{192\pi^3}F(x)\left[1 + \frac{\alpha}{2\pi}\left(\frac{25}{4}-\pi^2\right) \right]
    \label{eq:muon_lifetime_GF}
\end{equation}
where $F(x) = 1-8x-12x^2\ln x + 8x^3-x^4$ with $x=m_e^2/m_\mu^2$. Equation (\ref{eq:muon_lifetime_GF}) serves as the experimental definition of $G_F$ to $\mathcal{O}(\alpha)$. As a consequence, anything in the Standard Model EWRC calculation that is common to both the muon and nuclear $\beta$ decay can be absorbed into $G_F$ \cite{Sirlin1974}. In fact, standard methods result in the contribution of a number of divergent but process-independent integrals. When using an experimental determination of $G_F$, however, all other nuclear $\beta$ decay calculations are finite \cite{Sirlin1974, Marciano1975}. Taking into account higher-order corrections specific to the muon \cite{Sirlin2013}, the most precise value is found to be \cite{Tishchenko2013}
\begin{equation}
G_F/(\hbar c)^3 = 1.1663787(6) \times 10^{-5}\,\mathrm{GeV}^{-2}.
\label{eq:G_F_MuLan}
\end{equation}

Everything contained then in $RC$ of Eq. (\ref{eq:RC_def}) is specific to (nuclear) $\beta$ decays, relative to muon decay. In order to clearly denote the differences between $\Delta_R^{V, A}$ it is instructive to specify the precise origin of the pieces in the definition of $\Delta_R^V$. Taking the traditional breakdown as an example \cite{Marciano2006}, 
\begin{align}
    RC &= \frac{\alpha}{2\pi}\left[3\ln{\frac{m_W}{m_p}} + \overline{g}(E_0) + 6\overline{Q}_\beta \ln{\frac{m_W}{\Lambda}}\right. \nonumber \\
    &\left.  + 6(\overline{Q}_\beta - \overline{Q}_\mu)\ln{\frac{m_Z}{m_W}} + 2C_B  + 2C_\text{INT} + \mathcal{A}_g \right]
    \label{eq:inner_vector_blown_up}
    \\
    &+ \text{higher order}, \nonumber
\end{align}
where $\overline{Q}_\beta = 1/6$ is the average charge of up and down quarks, and $\overline{Q}_\mu = -1/2$ is the average charge of the $\mu^-$ and $\bar{\nu}_\mu$. The latter appears because we consider all effects relative to muon decay as mentioned above. The first two terms arise from low-energy photon exchange and contain an energy-average of Sirlin's famous $g$ function \cite{Sirlin1967}. The following two terms are asymptotic contributions from $\gamma W$ and $ZW$ box diagrams. Historically \cite{Marciano1986, Marciano2006}, the calculation is artificially divided in the loop momentum at some scale $m_p < \Lambda \ll m_W$. The benefit of this is that above this scale, the strong interaction is perturbative and gives rise to only small corrections. Below this scale, however, contributions from the axial part of the $\gamma W$ box are sensitive to physics at the nuclear scale and so are model-dependent. The final 3 terms are the main model-dependent parts of the calculation predominantly arising from the famous axial vector contribution to the Fermi decay rate. One receives contributions from the Born (elastic) term ($C_B$) at the nuclear scale, connects the two regimes through some interpolation function ($C_\text{INT}$) and adds small perturbative corrections from the deep inelastic scattering regime ($\mathcal{A}_g$). Recently it was shown \cite{Seng2019b}, however, that such a clear distinction in energy domains does not exist. We will come back to this in Sec. \ref{sec:gamma_W_box}.

\begin{figure}[ht]
    \centering
    \includegraphics[width=0.2\textwidth]{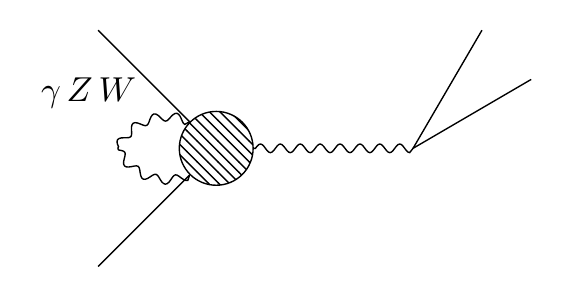}
    \includegraphics[width=0.2\textwidth]{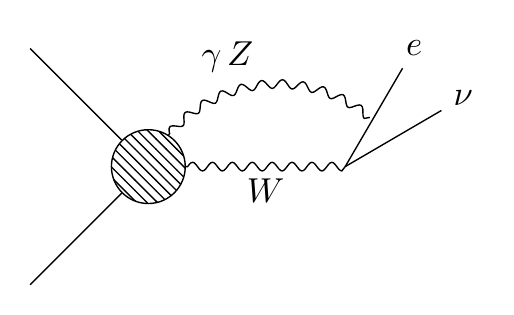}
    \caption{\label{fig:feynman_order_alpha}$\mathcal{O}(\alpha)$ radiative corrections that give rise to differences in vector and axial vector transitions.}
\end{figure}

Using Eq. (\ref{eq:inner_vector_blown_up}) it is now easy to see which terms are modified in the case of Gamow-Teller transitions. The first two terms do not depend on nuclear structure as they arise from the infrared-singular part of the $\gamma W$ diagram, which are known to be universal \cite{Sirlin1967}. Diagrams containing both virtual $W$ and $Z$ bosons can contribute only asymptotically to $\mathcal{O}(\alpha)$ because of the heavy boson propagators and $G_F \propto M_W^{-2} \ll 1$. In this regime, one essentially probes asymptotically free quarks and one obtains corrections proportional to the tree-level amplitude to lowest order. These give rise to the logarithmic enhancement factors of the third and fourth term in Eq. (\ref{eq:inner_vector_blown_up}) \cite{Sirlin1982}. As they are common for Fermi and Gamow-Teller transitions, they do not contribute to a difference in $\Delta_R^{V, A}$. Diagrams containing virtual photons, however, probe all scales, and will require the bulk of our attention. These remaining diagrams are shown in Fig. \ref{fig:feynman_order_alpha}.

\subsection{Common tools}

Following recent changes in CKM top-row unitarity results, a significant amount of research is being performed also in the $|V_{us}|$ sector \cite{Seng2019, Seng2020a}, some of which follow similar avenues as the ones taken here. Specifically, results based on current algebra are resurfacing, and will form the basis of our work. In the following sections, we discuss common elements to the calculation, and proceed with the evaluation of the vertex correction and $\gamma W$ box. We briefly summarize the other diagrams and their interaction with parts of the calculations of Fig. \ref{fig:feynman_order_alpha} in the appendix.

\subsubsection{Currents and commutation relations}
We follow the current algebra approach pioneered over 50 years ago \cite{Adler1968, Treiman1972, Sirlin1978}, and define the following quark currents
\begin{align}
    J^\mu_\gamma &= \frac{2}{3}\bar{u}\gamma^\mu u - \frac{1}{3}\bar{d}\gamma^\mu d \label{eq:J_gamma}\\
    J^\mu_W &= \bar{u}_L\gamma^\mu d_L \\
    J^\mu_Z &= \frac{1}{2}(\bar{u}_L\gamma^\mu u_L-\bar{d}_L\gamma^\mu d_L) \nonumber \\
    &- \frac{1}{3}\sin^2\theta_W (2\bar{u}\gamma^\mu u - \bar{d}\gamma^\mu d)
    \label{eq:J_Z}
\end{align}
where $\theta_W$ is the weak interaction angle and all quark fields obey canonical equal-time commutation relations (ETCR), ${\{\psi_a(t, \bm{x}), \psi_b^\dagger(t, \bm{y})\} = \delta_{ab}\delta^{(3)}(\bm{x}-\bm{y})}$. Using this, the ETCR for the currents of Eqs. (\ref{eq:J_gamma})-(\ref{eq:J_Z}) can directly be obtained and we find
\begin{subequations}
\begin{align}
    \left[J^0_\gamma(\bm{x}), J^\mu_W(0) \right] &=  J^\mu_W(\bm{x}) \delta^{(3)}(\bm{x}) \label{eq:ETCR_gamma_W} \\
    \left[J^0_W(\bm{x}), J^\mu_Z(0) \right] &= \cos^2\theta_W J^\mu_W(\bm{x}) \delta^{(3)}(\bm{x}) \\
    \left[J^0_W(\bm{x}), J^\mu_W(0) \right] &= -2\left[\sin^2\theta_WJ^\mu_\gamma(\bm{x})+J_Z^\mu(\bm{x})\right]\delta^{(3)}(\bm{x}) \label{eq:ETCR_W_W}
\end{align}
\end{subequations}
The appearance of the $\delta^{(3)}(\bm{x})$ factors will simplify matters significantly.

All the Feynman diagrams discussed in the following sections interfere linearly with the tree-level amplitude, which is simply
\begin{equation}
    \mathcal{M}_0 = -\frac{ig^2}{8}V_{ud}\frac{\bar{e}\gamma^\mu(1-\gamma^5)\nu}{q^2-M_W^2} \langle f | \bar{u}\gamma_\mu(1-\gamma^5)d| i \rangle
\end{equation}
as usual, with $g$ the $SU(2)_L$ gauge coupling and $|f, i\rangle$ are hadronic states which satisfy the strong interaction equation of motion. We define the lepton current
\begin{equation}
    L^\mu = \bar{e}_L\gamma^\mu\nu_L
    \label{eq:L_mu}
\end{equation}
for convenience and recognize 
\begin{equation}
    \frac{G_F}{\sqrt{2}} = \frac{g^2}{8M_W^2}
\end{equation}
for $q\ll M_W^2$, when making contact with the traditional Fermi four-point interaction of Eq. (\ref{eq:fermi_cc_interaction}).

\subsubsection{\texorpdfstring{$G$-parity}{G-parity} and first-class currents}
The strong interaction is symmetric under charge conjugation and isospin rotations. The combination of these, introduced by Lee and Yang \cite{Lee1956d}, is the so-called $G$-parity, defined as 
\begin{equation}
    G = C\exp \left(-i\pi T_2 \right)
\end{equation}
where $C$ is a charge conjugation operator and $T_2$ is the isospin projection along the $2$-axis. While the strong interaction is invariant under $G$-parity, both QED and the weak interaction are not. According to the scheme by Weinberg \cite{Weinberg1958}, all observed weak currents transform as first-class currents, meaning
\begin{align}
    G V_\mu G^{-1} &= V_\mu \\
    G A_\mu G^{-1} &= -A_\mu,
\end{align}
where $V_\mu$ transforms as a Lorentz vector and $A_\mu$ as an axial vector. In the absence of second-class currents (with the opposite behaviour) \cite{Wilkinson2000a, Triambak2017}, we can require the same thing from the radiative corrections. Specifically, all terms discussed in the following sections must individually transform as first-class currents. This is simply a way of quickly reducing the calculational load, as all terms which appear to transform as second-class vanish regardless in a full SM calculation.

\section{Electroweak vertex correction}
\label{sec:vertex_correction}
The first diagram under consideration is the vertex correction, where any of the three electroweak bosons couple directly to the vertex. A direct evaluation of its contribution is straightforward for a single nucleon, but generally more complex when moving into many-body systems. Regardless of the result, however, it must transform according to a $V-A$ structure to maintain Lorentz invariance when combined with $L^\mu$, Eq. (\ref{eq:L_mu}). Taking the photon as an example, we can write down an effective vertex operator, $\Gamma^\mu = \Gamma^\mu_0 + \delta \Gamma^\mu$, for a $J_i=1/2 \to J_f = 1/2$ transition between elementary fields
\begin{align}
    \delta\Gamma^\mu &= \frac{\alpha}{2\pi}\bar{u}_f \left[f_1(q^2)\gamma^\mu - i\frac{f_2(q^2)}{2M}\sigma^{\mu\nu}q_\nu + \frac{f_3(q^2)}{2M}q^\mu \right. \nonumber \\
    &\left.+ g_1(q^2)\gamma^\mu\gamma^5 - i\frac{g_2(q^2)}{2M}\sigma^{\mu\nu}q_\nu\gamma^5 + \frac{g_3(q^2)}{2M}q^\mu\gamma^5 \right]u_i,
\end{align}
where $f_i, g_i$ are dimensionless functions of $q = p_i-p_f$. All electroweak Standard Model currents which transform as a Lorentz vector are conserved, so that we can set $f_3$ to zero if initial and final states are on-shell. Further, since $g_2$ transforms as a second-class current, we can additionally set its influence to zero. This leaves \textit{a priori} four unknown form factors per virtual gauge boson. If one, as usual, neglects terms of $\mathcal{O}(q/M)$, the corrections do not depend on outgoing lepton momenta and contribute only to renormalize the effective coupling constants. In the following, we derive expressions for these form factors and discuss parts of their evaluation.

\subsection{Setting the stage I}
We follow Refs. \cite{Sirlin1978, Seng2020a} in using the on-mass-shell (OMS) perturbation formula. The latter states that for a general form factor
\begin{equation}
    F^\mu(p_f, p_i) = \langle f | \Gamma^\mu | i \rangle
\end{equation}
the modification to that form factor, $\delta F^\mu$, because of a change in the Lagrangian, $\delta \mathcal{L}$, can be written as
\begin{align}
    \delta F^\mu(p_f, p_i) &= \lim_{\bar{q}\to q} i T^\mu(\bar{q}, p_i, p_f) \nonumber \\
    &\equiv \lim_{\bar{q}\to q} \left[i \overline{T}^{\mu} -iB^\mu\right]
    \label{eq:delta_F}
\end{align}
where the tensor $T^{\mu}(\bar{q}, p_f, p_i)$ is
\begin{align}
T^{\mu}& = \int d^4ye^{i\bar{q}\cdot y} \langle p_f | T\{J_W^\mu(y)\delta \mathcal{L}(0)\} | p_i \rangle - B^\mu.
\label{eq:three_current_T}
\end{align}
Specifically, $B^\mu$ subtracts the contribution from the wavefunction renormalization of the outer legs of the vertex \cite{Seng2020a}, so that $\delta F^\mu$ is pole-free by construction\footnote{While the vertex correction is straightforward to obtain for neutron $\beta$ decay using elementary Feynman rules, the relations here are generally valid.}. For spin-$0$ systems this is
\begin{align}
    B^\mu(\bar{q}, p_i, p_f) &= -F^\mu(p_i-\bar{q}, p_i)\frac{i\delta m_f^2}{(p-\bar{q})^2-m_f^2} \nonumber \\
    &- F^\mu(p_f, p_f + \bar{q})\frac{i\delta m_i^2}{(p_f+\bar{q})^2-m_i^2}
    \label{eq:B_mu_Seng}
\end{align}
where $\delta m^2$ is the change in mass because of $\delta \mathcal{L}$, ${\delta m^2 = -\langle p | \delta\mathcal{L} | p \rangle}$, while for elementary spin-$1/2$ systems one writes \cite{Sirlin1978, SengPC},
\begin{align}
    B^\mu(\bar{q}, p_i, p_f) &= i\bar{u}(p_f)\left[ \frac{\delta m_f}{(\slashed{p_i}-\slashed{\bar{q}})-m_f}F^\mu(p_i-\bar{q}, p_i)\right. \nonumber \\
    &\left. + F^\mu(p_f, p_f + \bar{q})\frac{\delta m_i}{(\slashed{p_f}+\slashed{\bar{q}})-m_i}\right]u(p_i)
\end{align}
In the Standard Model, the loop bosons can be $a \in [\gamma, W, Z_0]$, and
\begin{align}
    \delta \mathcal{L}_a^\lambda{}_\lambda(0) &= \frac{C_a^2}{2(2\pi)^4} \int\frac{d^4k}{k^2-M_a^2} \nonumber \\
    &\times\int d^4x e^{ik \cdot x}  T\{J^\lambda_a(x)J_\lambda^a(0) \}
\end{align}
where $C_a$ are the electroweak coupling constants, i.e. $C_\gamma = e^2, C_W = g^2,$ and $C_{Z_0} = g^2+g^{'2}$, and $M_a$ is the physical boson mass. The analysis continues by coupling the OMS formula with the Ward-Takahashi identity (WTI) \cite{Sirlin1978, Seng2019b}. We start from
\begin{equation}
    iT^\mu = -\bar{q}_\nu \frac{\partial}{\partial \bar{q}_\mu}iT^\nu + \frac{\partial}{\partial \bar{q}_\mu}(i\bar{q}_\nu T^\nu)
    \label{eq:T_mu_identity}
\end{equation}
where, in particular, we are interested in the second term. We focus on $\overline{T}^\mu = \overline{T}^{\mu\lambda}{}_\lambda(\bar{q}, p_i, p_f)$ and perform a partial integration to arrive at
\begin{align}
    i\bar{q}_\nu \overline{T}^\nu &= -\frac{C_a}{2(2\pi)^4}  \int\frac{d^4k}{k^2-M_a^2}\int d^4x\int d^4ye^{i\bar{q}\cdot y}e^{ik\cdot x} \nonumber \\
&\times \partial_\nu\langle p_f | T\{J_W^\nu(y)J^\lambda_a(x)J^a_\lambda(0)\} | p_i \rangle.
\end{align}
The partial derivative of the time-ordered product of three currents obeys the identity
\begin{align}
    \frac{\partial}{\partial x^\nu} &T\left\{ J^\nu_W(x)J^\lambda_a(y)J_\lambda^a(0) \right\} = \nonumber \\ &T\biggl\{\partial_\nu J_W^\nu(x) J^\lambda_a(y)J_\lambda^a(0) \nonumber \\
    &+ \delta(x^0-y^0)\left[J^0_W(x), J^\lambda_a(y)\right]J_\lambda^a(0) \nonumber \\
    &+ \delta(x^0) \left[J^0_W(x), J^\lambda_a(0) \right]J_\lambda^a(y) \biggr\}.
    \label{eq:derivative_three_current}
\end{align}
For the currents defined here, the commutators were already derived in Eqs. (\ref{eq:ETCR_gamma_W})-(\ref{eq:ETCR_W_W}) and consist of a single current, a $c$-number and a Dirac delta. As a consequence, the vertex correction consists at least of a three-point correlation function and a two-point correlation function, corresponding to the first, and second and third terms, respectively. Using Eqs. (\ref{eq:delta_F}) and (\ref{eq:T_mu_identity})-(\ref{eq:derivative_three_current}), we can write the vertex correction matrix element as
\begin{align}
    \mathcal{M}_v^a &= \frac{g^2C_a}{4(2\pi)^4}V_{ud}\frac{L^\mu}{q^2-M_W^2} \lim_{\bar{q}\to q} \left[-\bar{q}_\nu \frac{\partial}{\partial \bar{q}^\mu}T^\nu_a \right.  \nonumber\\
    &\left. + \frac{\partial}{\partial \bar{q}^\mu}\biggl\{\mathcal{D}_a - \bar{q}_\nu B^\nu_a  + \mathcal{Z}^\lambda_{a}{}_\lambda\biggr\} \right]
    \label{eq:vertex_breakdown}
\end{align}
where
\begin{align}
    \mathcal{D}_a &= \int \frac{d^4k}{k^2-M_a^2}\int d^4y e^{i\bar{q}y}\int d^4x e^{ikx} \nonumber \\
    &\times \langle  p_f | T\left\{\partial_\mu J^\mu_W(y) J^\lambda_a(x)J_\lambda^a(0) \right\} | p_i \rangle
    \label{eq:D_a_3p}
\end{align}
is the three-point function correction, and
\begin{align}
    \mathcal{Z}_{a}^\lambda{}_\lambda(\bar{q}+k) &= \int \frac{d^4k}{k^2-M_a^2}\int d^4x e^{i(\bar{q}+k)x} \nonumber \\
    &\times\langle  p_f | T\left\{J^\lambda_b(x)J_\lambda^a(0) \right\} | p_i \rangle
\end{align}
is the two-point correlation function according to the ETCR, i.e. $\left[J^0_W(\bm{x}), J^\lambda_a(0) \right] \equiv J^\lambda_b(\bm{x})\delta^{(3)}(\bm{x})$.

Since $T^\mu_a$ is pole-free by construction, the contributions of the first term in Eq. (\ref{eq:vertex_breakdown}) is $\mathcal{O}(\alpha q) \sim \mathcal{O}(\alpha^2)$ since $q \sim 10^{-3}$. Setting $m_i = m_f$ in Eq. (\ref{eq:B_mu_Seng}), it is clear that the contribution of $B^\mu$ in Eq. (\ref{eq:vertex_breakdown}) is of order $q$. If one neglects terms of $\mathcal{O}(\alpha q)$, only contributions from $\mathcal{D}_a$ and $\mathcal{Z}^\lambda_{a}{}_\lambda$ remain. In all but the photonic case, $\mathcal{D}_a$ is insensitive to low $k^\mu$ due to the presence of the mass term in the heavy boson propagator. Specifically, since $M_{Z, W}^{-2} \propto G_F$ and the integrals are IR convergent, their contributions are $\mathcal{O}(G_F^2)$ and can safely be neglected. For the $W$ and $Z$ contributions then, only the asymptotic contributions for $k^\mu \to \infty$ contribute, specifically those coming from $x\sim y \sim 0$ and $y \sim 0$ for finite $x \neq 0$ \cite{Sirlin1978}. The former can be shown to be finite and of $\mathcal{O}(G_F^2)$, while the divergent contributions of the latter can be shown to cancel through the contribution of tadpole diagrams and order $\alpha$ counterterms \cite{Sirlin1978, Sirlin1982}. Finally then, only $\mathcal{D}_\gamma$ and $\mathcal{Z}^\lambda_{a}{}_\lambda$ give rise to finite contributions.

We can now move towards a simplification of the results. We recover the notation of Ref. \cite{Sirlin1978} by recognizing that
\begin{align}
    \frac{\partial}{\partial \bar{q}^\mu}\mathcal{Z}^\lambda_{a}{}_\lambda(\bar{q}+k) = \int\frac{d^4k}{k^2-M_a^2}\frac{\partial}{\partial k^\mu} T_{a}^\lambda{}_\lambda(\bar{q}+k)
    \label{eq:derivative_Z}
\end{align}
where
\begin{align}
    T^{\mu\nu}_\gamma(k) &= \int d^4x e^{ikx} \langle p_f | T\{J^\mu_\gamma(x)J^\nu_W(0)\} | p_i \rangle \label{eq:T_gamma} \\
    T^{\mu\nu}_Z(k) &= \int d^4x e^{ikx} \langle p_f | T\{J^\mu_Z(x)J^\nu_W(0)\} | p_i \rangle \label{eq:T_Z} \\
    T^{\mu\nu}_W(k) &= - \int d^4x e^{ikx}\biggl[ \sin^2\theta_W\langle p_f | T\{J^\mu_\gamma(x)J^\nu_W(0) \} | p_i \rangle  \nonumber \\
    &+ \langle p_f | T\{J_Z^\mu(x)J_W^\nu(0)\} | p_i \rangle \biggr] \label{eq:T_W}
\end{align}
are the Fourier transforms of two-current correlation functions.

We hold off on an evaluation of the two-point correlation functions until the next sections, but discuss some general features.
As before, both $T_Z^{\mu\nu}$ and $T_W^{\mu\nu}$ only depend on physics at and above the weak scale because of the heavy boson propagator. Their contributions should be considered together with additional graphs, and a detailed analysis shows that only finite terms survive that are common to Fermi and Gamow-Teller transitions \cite{Sirlin1978, Sirlin1982}. We provide a short summary in the Appendix. On the other hand, the photonic contribution, $T_\gamma^{\mu\nu}$, is sensitive to loop momenta of all scales and gives rise to non-asymptotic contributions. To neatly separate the latter from the asymptotic contributions we use a propagator trick introduced by Sirlin, where we write the photon propagator
\begin{equation}
    \frac{1}{k^2} = \frac{1}{k^2-m^2} + \frac{m^2}{m^2-k^2}\frac{1}{k^2}
    \label{eq:propagator_trick_Sirlin}
\end{equation}
where $m$ is an arbitrary mass scale. The first term can be interpreted as a massive photon with mass $m$, whereas the second term is the usual photon propagator with a Pauli-Villars (PV) regularization factor at $m$. If we set $m=M_W$, we recover the usual PV regularization factor in the old Fermi four-point theory \cite{Kinoshita1959}. Using this substitution and performing a partial integration of Eq. (\ref{eq:derivative_Z}) results in
\begin{align}
    &\int\frac{d^4k}{(2\pi)^4}\frac{1}{k^2}\frac{\partial}{\partial k_\mu} T_\gamma^{\lambda}{}_\lambda = \nonumber \\
    &-\int\frac{d^4k}{(2\pi)^4} T_\gamma^{\lambda}{}_\lambda \frac{\partial}{\partial k_\mu} \left(\frac{1}{k^2-M_W^2}+ \frac{M_W^2}{M_W^2-k^2}\frac{1}{k^2} \right)
    \label{eq:deriv_T_PI}
\end{align}
since the currents disappear at infinity. The first (`heavy photon') term combines with additional two-point correlation functions discussed in the Appendix and contributes only asymptotically through the Born term, i.e. common to both Fermi and Gamow-Teller transitions. The second term, on the other hand, contributes non-asymptotically and we write
\begin{align}
    \frac{\partial}{\partial k_\mu} \left( \frac{M_W^2}{M_W^2-k^2}\frac{1}{k^2} \right) &= \frac{2k^\mu}{k^2}\frac{M_W^2}{[M_W^2-k^2]^2} \nonumber \\
    &- \frac{2k^\mu}{k^4}\frac{M_W^2}{M_W^2-k^2}.
    \label{eq:derivative_PV_prop}
\end{align}
It is clear that the first term is $\mathcal{O}(G_F^2)$ for $k \ll M_W$ and vanishes for $k\to \infty$. The second is IR divergent and contains the so-called `convective' term, which is best combined with parts of the calculation of the $\gamma W$ box in Sec. \ref{sec:gamma_W_box}.

In summary, all terms arising from the vertex correction to $\mathcal{O}(\alpha)$ either vanish or are common to Fermi and Gamow-Teller transitions, with the exception of the photonic two-point and three-point functions. The former will be discussed in Sec. \ref{sec:gamma_W_box}, and we hold off on its evaluation. The latter, on the other hand, is unique to Gamow-Teller transitions and is discussed below.

\subsection{Three-point function evaluation}
\label{sec:3p_evaluation}
The photonic three-point function, $\mathcal{D}_\gamma$, depends on the divergence of the weak current as in Eq. (\ref{eq:D_a_3p}). For the vector transition case, i.e., the Fermi transition amplitude, the vector part of the weak interaction is conserved up to $\mathcal{O}(\alpha)$ (since isospin breaking correction can be thought of as order $\alpha$), so that $\mathcal{D}_\gamma^F = 0$ to the order of the calculation. In the general Gamow-Teller transition, however, this is not the case. We first look at its asymptotic behaviour, i.e. $k\to \infty$. While an operator product expansion (OPE) is straightforward, in this case we can equivalently use the Bjorken-Johnson-Low limit (BJL) \cite{Bjorken1964, Low1966}, with its three-current generalization given by Ref. \cite{Sirlin1968}. If for constant $\bar{q}^\mu$ and $\bm{k}$, $\mathcal{D}_\gamma \to 0$ for $k_0 \to \infty$, the BJL limit gives
\begin{align}
    \mathcal{D}_\gamma^{A} = &-\frac{1}{k_0^2}\int d^4y e^{i\bar{q}y}\int d^3x e^{i\bm{k}\cdot \bm{x}} \nonumber \\
    &\times\langle f | T \left\{\partial_\mu J^\mu(y) \left[\partial_0 J^\nu_a(\bm{x}), J_\nu^a(0) \right] \right\} | i \rangle \nonumber \\
    &+\frac{1}{k_0^2}\int d^3ye^{i\bm{q}\cdot \bm{y}}\int d^3x e^{i\bm{k}\cdot \bm{x}}\nonumber \\
    &\times\langle f | T \left\{\left[ \left[\partial_\mu J^\mu(\bm{y}), J^\nu_a(\bm{x})\right], J_\nu^a(0) \right] \right\} | i \rangle \nonumber \\
    & + \mathcal{O}\left(\frac{1}{k_0^3} \right)
\end{align}
where we added the superscript $A$ to denote the asymptotic piece. The $1/k_0$ term was set to zero since $\left[J^\nu_\gamma(\bm{x}), J_\nu^\gamma(0)\right] = 0$ under fairly general circumstances. In the asymptotic domain, the strong interaction is perturbative and quark fields are asymptotically free. To zeroth order in $\alpha_s$ then, one can use the canonical ETCR of Eqs. (\ref{eq:ETCR_gamma_W})-(\ref{eq:ETCR_W_W}) to evaluate the commutators. Following Ref. \cite{Sirlin1968}, the double commutator can be written as
\begin{align}
    &\left[ \left[\partial_\mu J^\mu(\bm{y}), J^\nu_a(\bm{x})\right], J_\nu^a(0) \right] = -\left[J^0(\bm{y}), \left[\partial_0 J^\nu_a(x), J_\nu^a(0) \right] \right] \nonumber \\
    &+ \delta^{(3)}(\bm{y}) \left[J_\mu(\bm{y}), \partial_0 J^\mu_a(\bm{x}) \right] - \delta^{(3)}(\bm{x}-\bm{y})\left[\partial_0J^\mu(\bm{y}), J_\mu^a(0) \right] \nonumber \\
    &+\frac{\partial}{\partial y^i} \left[\left[J^i(\bm{y}), J^\mu_a(\bm{x}) \right], J_\mu^a(0) \right]
\end{align}
All but the last are trivially evaluated using the ETCR of Eqs. (\ref{eq:ETCR_gamma_W})-(\ref{eq:ETCR_W_W}) and give rise to a $c$-number with $\delta^{(3)}(\bm{x}-\bm{y})\delta^{(3)}(\bm{x})$. The last double commutator can also be evaluated to give
\begin{align}
    \left[\left[J^i(\bm{y}), J^\mu_a(\bm{x}) \right], J_\mu^a(0) \right] &= 4 \delta^{(3)}(\bm{x}-\bm{y})\delta^{(3)}(\bm{x}) \nonumber \\
    &\times \bar{u}\gamma^0\gamma^i(1-\gamma^5)d
\end{align}
and the integral resolves to zero up to at least $\mathcal{O}(1/k_0^3)$. As a consequence, the asymptotic contribution to $\mathcal{D}_\gamma$ vanishes. This can also be intuitively understood thanks to the partially conserved axial current hypothesis. In the latter case the divergence of the axial current is non-zero only through the finite pion mass. Taking $Q^2 \to \infty$ means the divergence becomes negligible. Another way of understanding this is through chiral symmetry, where $\partial_\mu A^\mu$ vanishes above the chiral breaking scale $\Lambda_\chi$. Higher order QCD interactions modify this result only multiplicatively, and so the asymptotic contributions vanish to all orders in $\alpha_s$. The strong interaction becomes perturbative above the QCD scale, i.e. for $k \sim 1$ GeV. Since the asymptotic contributions vanish, the latter has no dependence on where we set this scale. 

Before moving on, we draw attention to an ambiguity in the evaluation of the time-ordered product in Eq. (\ref{eq:D_a_3p}), courtesy of Refs. \cite{Seng2019, SengPC}. Since the time-ordered product is not uniquely defined (i.e. Lorentz invariance requires the presence of a general $\delta(t)C(t)$), the derivative operator in $\partial_\mu J^\mu$ causes a problem. Specifically, using covariant perturbation theory this would translate into $\partial_\mu J^\mu \to q_\mu J^\mu$ where $q_\mu$ picks up off-shell momenta, and results in inconsistent behaviour with respect to the WTI+OMS approach described above. A way forward is to insert a complete set of on-shell states, and using an equation of motion $\partial_\mu J^\mu(x) = s(x)$ to make the substitution $\langle p_f | \partial_\mu J^\mu(x) | p_i \rangle = \langle p_f | s(x) | p_i \rangle$. We will use this property and in the discussion below use $\partial_\mu J^\mu_W = \partial_\mu A^\mu$ only schematically.

With this out of the way, we consider the Born channel as the low-energy contribution and we set
\begin{equation}
    \mathcal{D}_\gamma \approx \mathcal{D}_{\gamma}^\mathrm{Born}.
    \label{eq:D_gamma_approx_Born}
\end{equation}
It is important to keep in mind that the wave function renormalization contributions are subtracted by $B^\mu$ from the definition in Eq. (\ref{eq:three_current_T}). Further, because $\mathcal{D}_\gamma$ transforms like a pseudoscalar, it should be odd under $G$-parity. Given that the axial part of $J_W^\mu$ is odd, and the isoscalar (isovector) parts of $J^\mu_\gamma$ are odd (even) under $G$-parity, the double photonic current can only consist of $SS$ or $VV$ terms with no $SV$ iso-crossterms. This limits the number of contributing terms considerably.

We assume the coupling to the photon field as usual, with the Born response in the isospin formalism as
\begin{align}
    \mathcal{L}^I_{\gamma NN} &= i e \mathcal{A}^\mu_\gamma \bar{N} \left[F_1^I \gamma_\mu - i \frac{F_2^I}{2M}\sigma_{\mu \nu} \partial^\nu \right]T^{I} N  \nonumber \\
    &\equiv ie \mathcal{A}^\mu_\gamma \bar{N}\Gamma_\mu^I N
    \label{eq:EM_nucleon_vertex}
\end{align}
where $\mathcal{A}_\gamma^\mu$ is the photon field and $I$ can be either $0$ or $1$ for isoscalar and isovector contributions, respectively. The form factors are $F_1^1(0) = 1$, $F_2^1(0) = 3.706$, $F_1^0(0) = 1$ and $F_2^0(0) = -0.12$, and the isospin Pauli matrices are $T^1 = \tau^z$ and $T^0 = I_2$. The weak interaction elastic response for a nucleon is
\begin{align}
    W^\mu(p_2, p_1)& = \bar{N} \left\{g_V \gamma^\mu - i \frac{g_M}{2M}\sigma^{\mu\nu}q_\nu + \frac{g_S}{2M}q^\mu \right. \nonumber \\
    &\left. +g_A \gamma^\mu\gamma^5 - i\frac{g_T}{2M}\sigma^{\mu\nu}q_\nu\gamma^5 + \frac{g_P}{2M}q^\mu\gamma^5\right\} T^{\pm}N,
    \label{eq:current_decomp_nucleon}
\end{align}
where all $g_i$ are a function of $q = p_i-p_f$ and $T^\pm$ is the isospin ladder operator and $g_M = \kappa_p-\kappa_n = 3.706$ is the isovector magnetic moment using the conserved vector current hypothesis. The latter also forces $g_S = 0$. Assuming no second-class current exists \cite{Wilkinson2000a}, this additionally forces $g_T = 0$.

The Born contribution to $\mathcal{D}_\gamma$ is then
\begin{align}
    \mathcal{D}_\gamma^\mathrm{Born} &= \int \frac{d^4k}{k^2}\frac{\Lambda^2}{\Lambda^2-k^2}\bar{N}(p_f) \nonumber \\
    &\times \biggl[\Gamma^\lambda_I \frac{\slashed{p}_f-\slashed{k}+M}{k^2-2p_f\cdot k}\partial_\mu A^\mu \frac{\slashed{p}_i-\slashed{k}+M}{k^2-2p_i\cdot k}\Gamma_\lambda^I\biggr]N(p_i)
    \label{eq:D_gamma_Born}
\end{align}
where we have included the Pauli-Villars regularization factor at some scale $M \ll \Lambda \ll M_W$. Following the discussion above, after insertion of a complete set of on-shell states the transition depends on $\langle p' | \partial_\mu A^\mu | p \rangle$, with $A^\mu$ the axial vector part of $J_W^\mu$. Using the Dirac equation (thereby using on-shell nucleons), we can write
\begin{align}
    \langle p' | \partial_\mu A^\mu | p \rangle &= iq_\mu \langle p' | A^\mu | p \rangle \\
    &= i \left\{ 2M g_A(q^2) + \frac{q^2}{2M}g_P(q^2)\right\} \nonumber \\
    &\times[\bar{N}' \gamma^5 T^{\pm} N]
    \label{eq:div_A}
\end{align}
with $M$ the nucleon mass, and we used the decomposition of Eq. (\ref{eq:current_decomp_nucleon}) in the second line. Another way of estimating its impact is through the use of the PCAC hypothesis assuming pion-pole dominance. Specifically, we identify the divergence of the axial current with the pion field, and assume this to be equally valid near zero momentum transfer appropriate for $\beta$ decay rather than at $q^2 = m_\pi^2$ when taking $m_\pi \to 0$. In this case
\begin{align}
    \langle p' | \partial_\mu A^\mu | p \rangle &= i 2g_{\pi NN}F_{\pi NN}(q^2)f_\pi [\bar{N}' \gamma^5 T^{\pm} N]
    \label{eq:div_A_pcac}
\end{align}
where $f_\pi \approx 93$ MeV, $F(q^2=m_\pi^2) = 1$ and $g_{\pi NN}$ is the physical pion-nucleon coupling constant. Following through on PCAC and using the Goldberger-Treiman relationship we can additionally write \begin{equation}
    g_P(0) \simeq \frac{(2M)^2}{m_\pi^2}g_A(0) \approx -230
    \label{eq:gP_goldberger_treiman}
\end{equation}
so that Eq. (\ref{eq:div_A}) becomes
\begin{equation}
    \langle p' | \partial_\mu A^\mu | p \rangle \approx 2g_AM \left(1+\frac{q^2}{m_\pi^2}\right)[\bar{N}' \gamma^5 T^{\pm} N]
    \label{eq:div_a_gt}
\end{equation}
Returning to Eq. (\ref{eq:D_gamma_Born}), we assume $k \lesssim M$ due to the influence of the nucleonic form factors, $g_i(q^2)$, and evaluate in the center of mass frame of the initial state, i.e. $p_i = (M, \bm{0}) \approx p_f$. This simplifies matters greatly and we find
\begin{align}
    D_\gamma^\mathrm{Born} &= \int \frac{d^4k}{k^2}\frac{\Lambda^2}{\Lambda^2-k^2} (F_1^I)^2 \frac{\bar{N}(p_f)T^I\partial_\mu A^\mu T^IN(p_i)}{k_0^2 + i\epsilon}
\end{align}
when neglecting $\mathcal{O}(q/M)$ terms. We have not yet specified the isospin structure. The isoscalar nucleonic matrix element is given by, e.g., Eq. (\ref{eq:div_a_gt}) and gives a finite contribution when integration over $k$. Looking at the isospin structure of the isovector component, however, we have $T^1T^{\pm}T^1 = -T^{\pm}$ from properties of the Pauli matrices. We find then
\begin{align}
    D_\gamma^\mathrm{Born} &= 2g_AM \left(1+\frac{q^2}{m_\pi^2}\right)[\bar{N}' \gamma^5 T^{\pm} N] \nonumber \\
    &\times \int \frac{d^4k}{k^2} \biggl[(F_1^0)^2-(F_1^1)^2\biggr] \frac{\Lambda^2}{\Lambda^2-k^2} \frac{1}{k_0^2 + i\epsilon}.
    \label{eq:D_gamma_cancellation_scalar_vector}
\end{align}
In the isospin limit, for the nucleon $F_1^0(0) = F_1^1(0)$ and differences are small for $k \lesssim M$ \cite{Ye2018}. In this case, the Born contribution vanishes and so
\begin{equation}
    \mathcal{D}_\gamma \approx D_\gamma^\mathrm{Born} \approx 0.
\end{equation}
Therefore, to $\mathcal{O}(\alpha)$ the three-point function contribution to the vertex corrections is the same for Fermi and Gamow-Teller transitions. We note that this is only valid up to isospin breaking corrections, where the latter changes the commutator relations of $T^{\pm, z}$ operators, and introduces differences in $F_1^I$. We assume that these corrections are small (percent-level), and continue.

\section{Electroweak box diagrams}
\label{sec:gamma_W_box}
We arrive to the so-called box diagrams, with the exchange of a virtual photon or $Z$ boson between the initial or final state and the outgoing lepton as shown in Fig. \ref{fig:feynman_order_alpha}. As before, the $ZW$ box is insensitive to low-energy physics to $\mathcal{O}(G_F)$ because of the double heavy boson propagator. For $k \geq M_W$ the diagrams correspond only to a modification proportional to the tree-level amplitude \cite{Sirlin1978}, which we summarize in the appendix. The $\gamma W$ box diagram, on the other hand, is sensitive to effectively all scales, from $k\sim m_e$ to $k \gg M_W$. In the case of Fermi transitions, it contains the only remaining model dependence and is responsible for the theory uncertainty on the inner radiative correction \cite{Seng2018, Seng2019b, Czarnecki2019}. We will now discuss the $\gamma W$ box for Gamow-Teller transitions, where things become slightly more complex due to the non-conservation of the weak axial vector current.

\subsection{Setting the stage II}
The $\gamma W$ box matrix element is typically written as
\begin{align}
    \mathcal{M}_{\gamma W} &= -\frac{e^2g^2}{16}V_{ud} \int \frac{d^4k}{(2\pi)^4} \frac{1}{k^2[k^2 - 2l\cdot k][k^2-M_W^2]}\nonumber \\
    & \times \bar{e}(2l^\mu - \gamma^\mu\slashed{k})\gamma^\nu (1-\gamma^5)\nu T_{\mu \nu}^{\gamma W}
    \label{eq:gamma_W_box_def}
\end{align}
where $k$ is the internal loop momentum, $l$ is the external electron momentum and $T_{\mu \nu}^{\gamma W}$ is the so-called generalized Compton tensor of Eq. (\ref{eq:T_gamma}). In order to proceed, we use the well-known property of $\gamma$ matrices,
\begin{equation}
    \gamma^\mu\gamma^\lambda\gamma^\nu = g^{\mu\lambda}\gamma^\nu - g^{\mu\nu}\gamma^\lambda + g^{\lambda\nu}\gamma^\mu - i\epsilon^{\mu\lambda\nu\alpha}\gamma_\alpha \gamma^5,
    \label{eq:triple_gamma_reduction}
\end{equation}
to reduce the triple product of gamma matrices and we find
\begin{align}
    &\mathcal{M}_{\gamma W} = -\sqrt{2}\pi \alpha G_FV_{ud} \int \frac{d^4k}{(2\pi)^4} \frac{M_W^2}{k^2[k^2 - 2l\cdot k][k^2-M_W^2]}\nonumber \\
    & \times \biggl\{2l^\mu L^\nu - L^\nu k^\mu - L^\mu k^\nu + g^{\mu\nu}L^\lambda k_\lambda - i \epsilon^{\mu \lambda \nu \alpha} k_\lambda L_\alpha  \biggr\} \nonumber \\
    &\times T_{\mu \nu}^{\gamma W},
    \label{eq:gamma_W_box_init}
\end{align}
where we used $e^2 = 4\pi \alpha$ and $\epsilon^{\mu\nu\rho\sigma}$ is the completely asymmetric tensor with $\epsilon^{0123} = 1$. Following the ETCR of Eqs. (\ref{eq:ETCR_gamma_W})-(\ref{eq:ETCR_W_W}) one can construct two different WTI. The first of these is
\begin{equation}
    k^\mu T_{\mu\nu}^{\gamma W} = i \langle p_f | J_\nu^W(0) | p_i \rangle
    \label{eq:WTI_qed}
\end{equation}
where we used the conservation of the QED current, i.e. $\partial_\mu J^\mu_{\gamma} = 0$, while the second is
\begin{align}
    k^\nu T_{\mu\nu}^{\gamma W} &= i \langle p_f | J_\mu^W | p_i \rangle + q^\nu T_{\mu \nu}^{\gamma W} \nonumber \\
    &+ i\int d^4x e^{i (k - q) \cdot x} \langle p_f | T\{\partial^\nu J_\nu^W (x) J_\mu^{\gamma}(0) \} | p_i \rangle.
    \label{eq:WTI_weak}
\end{align}
For the remainder we drop the $\gamma W$ superscript on $T_{\mu\nu}$. Using the WTI, Eq. (\ref{eq:gamma_W_box_init}) reduces to
\begin{align}
    \mathcal{M}_{\gamma W} &= -\sqrt{2}\pi \alpha G_FV_{ud} \int \frac{d^4k}{(2\pi)^4} \frac{M_W^2}{k^2[k^2 - 2l\cdot k][k^2-M_W^2]}\nonumber \\
    & \times \biggl\{\text{TL} + 2l^\mu L^\nu T_{\mu \nu} - q^\nu L^\mu T_{\mu \nu} \nonumber \\
    &+ k_\nu L^\nu T^\lambda{}_\lambda - \mathcal{D}_\mu L^\mu + i \epsilon^{\mu \lambda \nu \alpha} k_\lambda L_\alpha T_{\mu \nu} \biggr\}
    \label{eq:gamma_W_box_WTI}
\end{align}
where `TL' stands for tree-level and $\mathcal{D}_\mu$ depends on the divergence of the weak current
\begin{equation}
    \mathcal{D}_\mu^\gamma = i\int d^4x e^{i (k - q) \cdot x} \langle p_f | T\{\partial^\nu J_\nu^W (x) J_\mu^{\gamma}(0) \} | p_i \rangle,
    \label{eq:D_2pt}
\end{equation}
in analogy with the three-point function correction of Eq. (\ref{eq:D_a_3p})\footnote{An equivalent expression is found in another recent work \cite{Seng2020a}.}.

Terms proportional to the tree-level amplitude are shared between Fermi and Gamow-Teller transitions and do not contribute to a difference in $\Delta_R^{V, A}$ of Eq. (\ref{eq:gA_def_QCD}). The second term in Eq. (\ref{eq:gamma_W_box_WTI}) is part of the infrared divergent contribution as categorized in Ref. \cite{Sirlin1967} and becomes part of the common so-called outer corrections which depend on the electron momentum but is independent of the strong interaction. Neglecting effects of $\mathcal{O}(\alpha q/M)$ as we have done before, the third term in the second line of Eq. (\ref{eq:gamma_W_box_WTI}) can equally be set to zero, and only the last line in Eq. (\ref{eq:gamma_W_box_WTI}) remains. Of these three terms, the first cancels to the order of the calculation with a contribution of the photonic vertex correction of Eqs. (\ref{eq:vertex_breakdown}), (\ref{eq:derivative_Z}) and (\ref{eq:derivative_PV_prop}). Specifically, we can rewrite the denominator of the $\beta$ particle propagator of Eq. (\ref{eq:gamma_W_box_WTI}) as
\begin{equation}
    \left(k^2-2l\cdot k\right)^{-1} = \frac{1}{k^2} + \frac{2 l\cdot k}{k^2(k^2-2 l\cdot k)}
    \label{eq:fermion_propagator_trick}
\end{equation}
so that the photonic vertex contribution of Eq. (\ref{eq:derivative_PV_prop}) cancels exactly with the first term. The second term, on the other hand, vanishes for $k\to \infty$ (since $T_{\mu\nu} \sim 1/k$) but is infrared-divergent and contributes to the outer corrections. This was noted already long ago \cite{Sirlin1978} and reiterated in another recent work \cite{Seng2020a}. Thereby both two-point and three-point functions of the vertex correction in the previous section have been dealt with. In Sec. \ref{sec:cancellation} we show that this cancellation is not taken into account in the traditional $\beta$ decay calculations leading to important discrepancies.

Finally, this leaves the contribution of the divergence of the weak current, $\mathcal{D}_\mu$, and the parity-odd part of $T^{\mu\nu}$. For a vector transition the former vanishes due to the conservation of the weak vector current, whereas the non-zero divergence of the weak axial current contributes \textit{a priori} to the Gamow-Teller transition. For vector transitions, the parity-odd contribution is the only remaining model dependence in the evaluation of $\Delta_R^V$, i.e. the famous axial input to the $\gamma W$ box \cite{Abers1967, Sirlin1978}, which has inspired research for well over half a century \cite{Abers1968, Marciano1984, Marciano1986, Czarnecki2004, Marciano2006, Sirlin2013, Seng2018, Seng2019b, Czarnecki2019}. Analogously, for Gamow-Teller transitions the parity-odd contribution arises from the vector part of $T^{\mu\nu}$ to the axial amplitude. Although some differences arise, we will see that their treatment is very similar when the dust has settled.

In the case of a vector transition the generalized forward Compton tensor is
\begin{equation}
    A^{\mu\nu} = i\int d^4x e^{ikx} \langle p_f s_f | T\{J_\gamma^\mu(x)A^\nu(0)\} | p_i s_i \rangle
    \label{eq:T_F}
\end{equation}
where $A^\nu$ is the axial vector component of $J^\nu_W$ as before. For a Fermi transition there is no angular momentum dependence besides the requirement that initial and final spins are equal. Further, since the parity-odd term does not contribute at $k\sim m_e$, we can neglect the outgoing lepton momentum and set $p_i=p_f=p$ and $k_e\to 0$. Therefore, using Lorentz invariance, one can decompose the forward $T^{\mu\nu}$ tensor for Fermi transitions into its constituent structure functions after summing over all spins. The axial current, however, is not conserved and the former then requires 14 different structure functions \cite{Ji1993, Maul1997}. Because of the contraction with the Levi-Civita tensor in Eq. (\ref{eq:gamma_W_box_WTI}) and the absence of spin dependence for a Fermi transition, however, only a single structure function survives
\begin{equation}
    A^{\mu\nu} \stackrel{\text{asy}}{\longrightarrow} i\frac{\epsilon^{\mu \nu \alpha \beta}p_\alpha k_\beta}{2(p\cdot k)}\mathcal{A}_3(\nu, Q^2),
    \label{eq:expansion_structure_functions}
\end{equation}
with $\nu = p\cdot k / M$ the energy transfer and $Q^2 = -k^2$ the photon virtuality. Following the usual notation for the photonic box diagram contribution, this allows one to write\cite{Seng2019b, Marciano2006}
\begin{equation}
    \mathcal{M}_0+\mathcal{M}_{\gamma W} = \sqrt{2}g_VG_FV_{ud}(1+\Box_{\gamma W}^{VA})p_\mu L^\mu
\end{equation}
where
\begin{equation}
    \Box_{\gamma W}^{VA} = \frac{4\pi \alpha}{g_V(0)} \mathrm{Re} \int \frac{d^4k}{(2\pi)^4}\frac{M_W^2}{M_W^2+Q^2}\frac{Q^2+\nu^2}{Q^4}\frac{\mathcal{A}_3(\nu, Q^2)}{M\nu}.
    \label{eq:box_VA_T3}
\end{equation}

Analogous to Eq. (\ref{eq:T_F}), the Gamow-Teller transition receives contributions only from
\begin{equation}
    V^{\mu\nu} = i\int d^4x e^{ikx} \langle p_f s_f | T\{J_\gamma^\mu(x)V^\nu(0)\} | p_i s_i \rangle
    \label{eq:V_munu}
\end{equation}
with $V^\nu$ the weak vector current. Because the latter is conserved, however, an expansion like Eq. (\ref{eq:expansion_structure_functions}) is simplified and only 7 structure are required\footnote{Because of the spin independence of the Fermi matrix element and the contraction with the Levi-Civita tensor, however, the simplification is merely conceptual.} \cite{Blumlein1997}. If we once more write only terms that survive the contraction with the Levi-Civita tensor, we write \cite{Ji1993}
\begin{align}
    V^{\mu\nu} &\stackrel{\text{asy}}{\longrightarrow} i\epsilon^{\mu\nu\alpha\beta} \frac{k_\alpha p_\beta}{2 (p\cdot k)}\mathcal{V}_3(\nu, Q^2) \nonumber \\ 
    &+i\epsilon^{\mu\nu\alpha\beta}\frac{k _\alpha}{p\cdot k}\biggl[S_\beta \mathcal{G}_1(\nu, Q^2) \nonumber \\
    &+\left. \left(S_\beta-p_\beta\frac{S\cdot k}{p\cdot k}\right) \mathcal{G}_2(\nu, Q^2)\right]
    \label{eq:expansion_structure_functions_axial_asy}
\end{align}
where $S_\beta$ is the polarization four-vector. The latter is equal to $S_\beta = (0, \bm{S})$ in the rest frame of the initial state and normalized as $S^2 = -M^2$ \cite{Zyla2020}. Similarly as above, we define
\begin{equation}
    \mathcal{M}_0+\mathcal{M}_{\gamma W} = \sqrt{2}g_AG_FV_{ud}(1+\Box_{\gamma W}^{VV}) S_\mu L^\mu
\end{equation}
where
\begin{align}
    \Box_{\gamma W}^{VV}= -\frac{4\pi\alpha}{g_A(0)}\mathrm{Re} &\int \frac{d^4k}{(2\pi)^4}\frac{M_W^2}{k^4[k^2-M_W^2]M\nu} \nonumber \\
    &\times \left[\mathcal{G}_1(\nu, Q^2)\frac{2k^2+\nu^2}{3}+\mathcal{G}_2k^2 \right].
\end{align}
This equation can be used as the starting point for a dispersion relation analysis, which lies beyond the scope of this manuscript.

In summary, the total difference in contributions for Fermi to Gamow-Teller transitions from the $\gamma W$ box diagram is then
\begin{equation}
    \Delta_R^A-\Delta_R^V = 2(\Box_{\gamma W}^\mathcal{D}+\Box_{\gamma W}^{VV}-\Box_{\gamma W}^{VA})
\end{equation}
with $\Box_{\gamma W}^\mathcal{D}$ the contribution of the $\mathcal{D}_\mu$ term in Eq. (\ref{eq:D_2pt}).

\subsection{Axial divergence}
Here, we consider the contribution of the $\mathcal{D}_\mu$ term in Eqs. (\ref{eq:gamma_W_box_WTI}) and (\ref{eq:D_2pt}). Since the weak vector current is conserved it vanishes for a pure Fermi transition and contributes \textit{a priori} to a Gamow-Teller decay. We will discuss its asymptotic and Born contributions separately.

In Sec. \ref{sec:3p_evaluation} we argued that the partial conservation of the axial current meant it did not lead to UV divergences. This can once again be shown using an operator product expansion or the BJL limit. The result will in this case be identical, and we write to $\mathcal{O}(1/k_0)$
\begin{equation}
    \lim_{k_0\to \infty} D_\nu = \frac{i}{k_0} \int d^3x e^{-i\bm{k}\cdot \bm{x}} \langle p_f |[\partial^\mu J_\mu(\bm{x}), J_\nu^\gamma(0)] | p_i \rangle.
    \label{eq:D_2pt_BJL}
\end{equation}
We can evaluate the commutator explicitly using the ETCR of Eqs. (\ref{eq:ETCR_gamma_W})-(\ref{eq:ETCR_W_W}). Because the Standard Model is a local theory, however, the commutator is proportional to $\delta^{(3)}(\bm{x})$, and it is clear from Eq. (\ref{eq:D_2pt_BJL}) that the asymptotic contribution of $D_\mu$ vanishes. This coincides with our initial reasoning based on the partial conservation of the axial current or chiral invariance.

Since the asymptotic contribution vanishes, we can analogously to Sec. \ref{sec:3p_evaluation} define some separation energy scale $\Lambda \sim$ few GeV above which the strong interaction can be considered perturbative and we may apply the BJL limit. Below this scale we consider only the Born amplitude, so that like in Eq. (\ref{eq:D_gamma_approx_Born}) we write
\begin{equation}
    \mathcal{D}_\mu \approx \mathcal{D}_\mu^\mathrm{Born}.
\end{equation}
Like our discussion above for the three-point contribution, $\mathcal{D}_\gamma$, we use the divergence, $\partial_\mu A^\mu$, only schematically and instead use, e.g., the PCAC hypothesis. The Born amplitude then is
\begin{align}
    \mathcal{D}_\mu^\mathrm{Born} &=\bar{N}\left[ \Gamma^\mu\frac{\slashed{p}_f+\slashed{k}+M}{k^2+2p_f\cdot k + i\epsilon} \partial_\nu A^\nu\right. \nonumber \\
    &\left.+  \partial_\nu A^\nu\frac{\slashed{p}_i-\slashed{k}+M}{k^2-2p_i \cdot k + i \epsilon}\Gamma^\mu\right]N,
    \label{eq:D_2p_Born}
\end{align}
with the notation of Sec. \ref{sec:3p_evaluation}. In the Born amplitude the form factors decrease strongly with increasing $k$, so that we may neglect $k^2$ against $p\cdot k$, and set the latter equal to $Mk_0$ in the initial rest frame with impunity. The error we make with this is $\mathcal{O}(m_e/M)$ and is small. We then find, keeping only the $+i\epsilon$ parts
\begin{equation}
    \frac{1}{k^2+2 p_f\cdot k + i\epsilon} + \frac{1}{k^2-2p_i \cdot k + i\epsilon} \approx - i\pi \frac{\delta(k_0)}{M}.
    \label{eq:propagator_res_trick}
\end{equation}
Finally, when invoking $G$-parity it is obvious that only the isovector part of $J_\gamma$ can contribute to $D_\mu$ since $\bar{N}\partial_\nu A^\nu N$ transforms like a pseudoscalar. Writing only the monopole term for clarity
\begin{align}
    \mathcal{D}_\mu^\mathrm{Born} \approx -i\pi\delta(k_0) F_1^1 \bar{N} \left[T^z \partial_\nu A^\nu + \partial_\nu A^\nu T^z \right] N,
\end{align}
where it is important to note that $\partial_\nu A^\nu \propto T^{\pm}$ as discussed above. Using the anti-commutation properties of the Pauli matrices, i.e. $\{\sigma_a, \sigma_b\} = 2\delta_{ab}I_2$, we see that the result vanishes since $\{T^z, T^{\pm}\} = 0$, and so $\mathcal{D}_\mu^\mathrm{Born} = 0$. Analogous to Sec. \ref{sec:3p_evaluation}, we find that both the asymptotic and finite parts vanish, and so 
\begin{equation}
    \Box_{\gamma W}^\mathcal{D} \approx 0.
\end{equation}
This leaves only the polarized parity-odd contribution, analogous to Fermi transitions.

\subsection{Parity-odd amplitude}
\label{sec:parity_odd}
With all other terms in Eq. (\ref{eq:gamma_W_box_WTI}) either common to Fermi and Gamow-Teller transitions or the parts specific to the latter found to vanish, only the parity-odd term remains. We will be somewhat more careful here and consider not only the asymptotic and Born contributions, but also the intermediate energy regime and perturbative QCD corrections. We simplify the notation of the final term in Eq. (\ref{eq:gamma_W_box_WTI}) by introducing a general function $F^{A,V}(Q^2)$
\begin{align}
    \Box_{\gamma W} = \frac{\alpha}{2\pi} \int_0^\infty dQ^2\frac{M_W^2}{Q^2+M_W^2} F^{A,V}(Q^2)
    \label{eq:box_Q2_general}
\end{align}
where we Wick rotated the momentum integral and adopted a notation similar to Ref. \cite{Marciano2006}. Now, $F^A(Q^2)$ denotes the contribution to Gamow-Teller transitions, and $F^V(Q^2)$ that of Fermi transitions. We first introduce the more straightforward elements and build in complexity to arrive at a consistent description.

\subsubsection{Born contribution}
We start with the most straightforward part of the amplitude, which is the Born contribution for low $Q^2$. The Born amplitude of $T^{\mu\nu}$ in Eq. (\ref{eq:gamma_W_box_WTI}) can be written in the isospin formalism as
\begin{align}
    T^{\mu \nu}_\text{Born} = \bar{N} &\left[ \Gamma^\mu_I\frac{\slashed{p}_f-\slashed{k}+M}{k^2-2p_f\cdot k + i\epsilon} W^\nu(p_f -k, p_i)\right. \nonumber \\
    &\left.+  W^\nu(p_f, p_i+k)\frac{\slashed{p}_i+\slashed{k}+M}{k^2+2p_i \cdot k + i \epsilon}\Gamma^\mu_I\right]N,
    \label{eq:T_munu_Born}
\end{align}
where $W^\nu(p_2, p_1)$ is the weak transition matrix element of Eq. (\ref{eq:current_decomp_nucleon}), and $\Gamma^\mu_I$ the electromagnetic vertex of Eq. (\ref{eq:EM_nucleon_vertex}) for isoscalar ($I=0$) or isovector ($I=1$) parts. We perform some reduction of $\gamma$ matrices for bookkeeping. The monopole terms are easy to treat, and the numerator in each fermion propagator can simply be replaced by $2p^\mu \pm \gamma^\mu \slashed{k}$, wheres the $F_2$ terms are somewhat more involved
\begin{align}
    &i\frac{F_2}{2M}\bar{N}\sigma^{\mu\nu}k_\nu (\slashed{p}_f+\slashed{k}+M) = -\frac{F_2}{2M}\bar{N}k_\nu (p^\nu \gamma^\mu - p^\mu \gamma^\nu \nonumber \\
    &- i\epsilon^{\sigma \mu \nu \rho}\gamma_\sigma p_\rho \gamma^5 + k^\nu\gamma^\mu - k^\mu \gamma^\nu + M\sigma^{\mu\nu}k_\nu).
\end{align}
The calculation is simplified by noting that the on-shell nucleons are highly non-relativistic, which means that any product of $\gamma$ matrices must have non-zero diagonal elements, lest the matrix element be suppressed by a relativistic factor $v/c$. Additionally, we can set $p^\mu_{i,f} \approx (M, \bm{0})$ in the center of mass frame. Finally, when combined with the lepton tensor $L_\mu$, one must have $\mu = 0$ for it to contribute to the Fermi box, whereas $\mu$ must be spacelike for Gamow-Teller. It is then straightforward to show that the Fermi amplitude receives contributions only from the main Gamow-Teller term, $g_A \gamma^\mu \gamma^5$, whereas the Gamow-Teller transition receives contributions from both the leading Fermi amplitude, $g_V\gamma^\mu$, and weak magnetism contribution, $g_M \sigma^{\mu\nu}k_\nu$. Specifically,
\begin{align}
    F^V_\mathrm{Born} &= \frac{1}{Q^2}\frac{|g_A|(F_1+F_2)}{g_V(0)}  \frac{1+2r}{(1+r)^2} \label{eq:F_V_Born} \\
    F^A_\mathrm{Born, LO} &= \frac{1}{Q^2}\frac{g_V(F_1+F_2/2)}{|g_A(0)|}\frac{5+4r}{3(1+r)^2} \\
    F^A_\mathrm{Born, WM} &=\frac{1}{Q^2}\frac{g_MF_1}{|g_A(0)|}\frac{5+4r}{6(1+r)^2}
\end{align}
where for the weak magnetism part only the monopole contributes up to $\mathcal{O}(1/M)$ and $r=\sqrt{1+4M^2/Q^2}$. We discuss the calculation in some more detail in Appendix \ref{app:born}.

So far, we have not explicitly mentioned the isospin structure of the electromagnetic interaction. While one can perform the calculations explicitly \cite{Towner1992}, we can invoke $G$-parity instead. Since all terms must be even (odd) for Fermi (Gamow-Teller) transitions, only the isoscalar part contributes to both. Therefore, we can replace $F_i$ everywhere by $F_i^{0}$, with the charges as defined in Sec. \ref{sec:3p_evaluation}. As a consequence, the magnetic interaction is strongly suppressed and it is mainly the monopole interaction that dominates.

Previously, the Born contribution has been treated in two ways with regards to its integration domain. In one \cite{Marciano2006, Czarnecki2019}, it is integrated only to the onset of perturbative QCD (pQCD) results, whereas in the other \cite{Seng2019b} all contributions up to infinity are included. We argue that the latter is consistent with our approach, as the pQCD results discussed below were originally derived far away from the elastic regime. When comparing to data, however, it is imperative to include also the elastic contribution at all scales in order to, e.g., determine higher-twist corrections \cite{Ji1993a, Deur2008}. And so, integrating out to $Q^2 \to \infty$ we find
\begin{align}
    \Box_F^\text{Born} &= 0.91(5) \frac{\alpha}{2\pi} \label{eq:box_F_Born}\\
    \Box_{GT}^\text{Born} &= [0.39(1) + 0.78(2)] \frac{\alpha}{2\pi}\\
    &= 1.17(2)\frac{\alpha}{2\pi}
    \label{eq:box_GT_Born}
\end{align}
where we have split up the leading order and weak magnetism induced effect, and the uncertainty arises from the form factors added in quadrature \cite{Seng2019b}. The uncertainty in the Gamow-Teller contribution is smaller because the vector form factors are known to higher accuracy. Our result for the Fermi contribution agrees exactly with Ref. \cite{Seng2019b}, as expected. It is interesting to note that $\Box_{GT}^\mathrm{Born}$ is dominated by the induced weak magnetism contribution rather than the leading-order term. The latter is reduced compared to the Fermi contribution due to the faster decrease in the vector form factor and the overall $g_V/g_A$ prefactor. The normalization with respect to $g_A(0)$ makes the overall axial correction substantially smaller than the raw $\gamma W$ box integral, which is almost 70$\%$ larger in the axial vector case relative to the vector transition.

\subsubsection{Deep inelastic scattering}

We continue by describing the asymptotic behaviour to zeroth order in $\alpha_s$. This can readily be obtained from the BJL limit or an OPE and we retain only the asymmetric tensor part to arrive at
\begin{equation}
    \lim_{k_0\to \infty} T^{\mu\nu}_\text{asy} = \frac{2\bar{Q}}{k^2} \epsilon^{\mu\nu\rho\sigma}k_\rho \langle p_f | J_\sigma^W(0) | p_i \rangle
    \label{eq:T_gamma_W_OPE}
\end{equation}
where $\bar{Q}$ is the average of the quark charges. In combination with the Levi-Civita tensor of Eq. (\ref{eq:gamma_W_box_WTI}) this results in
\begin{align}
    \lim_{k_0\to \infty} \epsilon^{\mu \nu \lambda \alpha} &k_\lambda L_\alpha T_{\mu \nu} = \frac{4\bar{Q}k^2}{k^2-2p\cdot k}\nonumber \\
    & \times\left(g^\alpha{}_\sigma-\frac{k_\sigma k^\alpha}{k^2} \right) \langle p_f | J_W^\sigma(0) | p_i \rangle L_\alpha,
    \label{eq:T_gamma_W_OPE_contraction}
\end{align}
as expected. Since this is once again proportional to the tree-level amplitude, it is common for Fermi and Gamow-Teller transitions and so does not contribute to a renormalization unique to $g_A$. In fact, as the leading behaviour of Eq. (\ref{eq:T_gamma_W_OPE_contraction}) is independent of $k$ in the UV, Eq. (\ref{eq:gamma_W_box_WTI}) gives rise to logarithmic enhancement factors $\bar{Q}\ln M_W$ when performing the $k$ integration, as mentioned in Sec. \ref{sec:outline} and various places in the literature \cite{Abers1968, Sirlin1982}.

The result of Eq. (\ref{eq:T_gamma_W_OPE}) is valid only to zeroth order in $\alpha_s$, above some scale $M \ll \Lambda \ll M_W$. In order to include higher-order QCD contributions in the perturbative ($Q^2 \gtrsim \Lambda$) regime, we follow the reasoning of Refs. \cite{Marciano2006, Seng2019b}. Specifically, Marciano and Sirlin \cite{Marciano2006} realized that the running of $T_{\mu\nu}$ can be related to that of the polarized Bjorken sum rule through a chiral transformation (see Appendix). Since the QCD Lagrangian is chirally symmetric above $\Lambda_{\chi} \sim 1$ GeV, this relation holds for deep inelastic scattering where $Q^2 \gg \Lambda_{\chi}$. The polarized Bjorken sum rule (PBjSR) is written in terms of the difference in Mellin moments of proton and neutron
\begin{align}
    \Gamma_1^{p-n}(Q^2) &= \int_0^{1}dx [g_1^p(x, Q^2)-g_1^n(x, Q^2)] \label{eq:PBjSR_def} \\
    &= \frac{|g_A|}{6}\left[1-\frac{\alpha_{g_1}(Q^2)}{\pi} \right]
    \label{eq:PBjSR_DIS}
\end{align}
where $x = Q^2/2M\nu$ is the Bjorken-$x$, $g_1^{p(n)}$ is the polarized structure function of the proton (neutron) and
\begin{equation}
    1-\frac{\alpha_{g_1}(Q^2)}{\pi} = \left[1 - \sum_{i=1}^N C_i^\mathrm{Bj} \left(\frac{\alpha_s}{\pi} \right)^i\right].
    \label{eq:alpha_g1}
\end{equation}
Corrections up to $\mathcal{O}(\alpha^4)$ are known in the $\overline{\text{MS}}$ scheme \cite{Larin1991, Baikov2010}, with $C_1^\mathrm{Bj} = 1$, $C_2^\mathrm{Bj}=\frac{55}{12}-\frac{1}{3}N_f$, $C_3^\mathrm{Bj} = 41.440-7.607N_f+0.177N_f^2$, and $C_4^\mathrm{Bj} = 479.4 - 123.4N_f + 7.697N_f^2 - 0.1037N^3_f$ where $N_f$ is the number of active flavours discussed below.

In Ref. \cite{Seng2019b} one also explored using isospin symmetry to relate $T_{\mu\nu}^{\gamma W}$ to (anti)neutrino-nucleon scattering. The argument can be summarized as follows: The optical theorem \& Schwarz reflection principle relates the forward amplitude of Eq. (\ref{eq:expansion_structure_functions}) to the analogous structure function, $F_3^{\gamma W}(\nu, Q^2)$, of the \textit{full} hadronic tensor via
\begin{equation}
    \mathrm{Dis}~ \mathcal{A}_3(\nu, Q^2) = 4 \pi F_3^{\gamma W}(\nu, Q^2)
    \label{eq:F3_T3_opt}
\end{equation}
where for unpolarized states
\begin{align}
    W^{\mu\nu}_{\gamma W} &= \frac{1}{4\pi}\sum_X(2\pi)^4 \delta^4(p+k-p_X)\langle p | J^\mu_\gamma | X \rangle \langle X | J^\nu_W | p \rangle \nonumber \\
    &= \ldots + \frac{i\epsilon^{\mu\nu\rho\sigma}p_\rho k_\sigma}{2(p\cdot k)}F_3^{\gamma W}(\nu, Q^2) 
\end{align}
with $X$ all possible intermediate states. The $F_3^{\gamma W}$ structure function of the weak axial vector and photonic current is not experimentally accessible, however, and one instead performs an isospin rotation $\gamma W \to WW$. Such a process is probed in charged current (anti)neutrino-nucleon scattering and which reveals $F_3^{\nu p}(\nu, Q^2)$ and $F_3^{\bar{\nu}p}(\nu, Q^2)$. The latter \textit{are} known experimentally, and $\alpha_s$ corrections are known in the deep inelastic scattering regime from the running of the Gross-Llewellyn Smith (GLS) sum rule \cite{Larin1991}
\begin{equation}
    \int_0^1dx [F_3^{\nu p}(x, Q^2) + F_3^{\bar{\nu}p}(x, Q^2)] = 3 \left[1-\frac{\alpha_{F_3}(Q^2)}{\pi}\right]
    \label{eq:GLS_def}
\end{equation}
with $x$ as above and 
\begin{equation}
    1 - \frac{\alpha_{F_3}(Q^2)}{\pi} = \left[1 - \sum_{i=1}^N C_i^\mathrm{GLS} \left(\frac{\alpha_s}{\pi} \right)^i\right] \label{eq:alpha_F3}
\end{equation}
writing only the leading twist result as before. Corrections are similarly available up to N$^4$LO \cite{Larin1991, Baikov2010, Baikov2012}, and largely the same as those for the PBjSR. Differences show up at $\mathcal{O}(\alpha_s^3)$ due to singlet (light-by-light) contributions and one finds $C_3^\mathrm{GLS} = 41.440 - 8.020N_f + 0.177N_f^2$ and $C_4^\mathrm{GLS} = 479.4 - 117.6N_f+7.464N_f^2-0.1037N_f^3$. 
With some foresight, we entertain both GLS and PBj sum rule treatments for the vector transition and write
\begin{equation}
    F^V_\mathrm{DIS}(Q^2) \approx \dfrac{1}{4Q^2}\left\{\begin{array}{cc}
        1 - \dfrac{\alpha_{F_3}(Q^2)}{\pi} & \mathrm{(GLS)} \\
        1 - \dfrac{\alpha_{g_1}(Q^2)}{\pi} & \mathrm{(PBj)}
    \end{array}  \right.
    \label{eq:F_V_DIS}
\end{equation}
Because of the large similarity between the two, however, we anticipate differences to be small.

In the case of the axial transition, the correspondence is much more transparent and the running of $T_{\mu\nu}$ can easily be related to that of the polarized Bjorken sum rule (see Appendix). Once again neglecting isospin breaking corrections, we can therefore write
\begin{equation}
    F^{A}_\mathrm{DIS}(Q^2) \approx \frac{1}{4Q^2}\left[1 - \frac{\alpha_{g_1}(Q^2)}{\pi}\right].
    \label{eq:F_A_DIS}
\end{equation}

Before moving on we briefly touch upon on the number of active flavours participating in the running, $N_f$. The pQCD corrections to the sum rules discussed above are derived in the limit of massless quarks, which implies $N_f=3$ at reasonably low $Q^2$ since charm and bottom are decoupled. Reference \cite{Czarnecki2019} takes into account these heavy quarks by incrementing $N_f$ when $Q^2$ exceeds some decoupling thresholds $m_c$ and $m_b$, causing discrete jumps in the $\alpha_{g_1}$ function. When taking into account also massive flavour corrections \cite{Blumlein2016}, however, this increment becomes quenched. In fact, when including these additional corrections the $N_f = 5$ result is reached only asymptotically for $Q^2\to \infty$, and the effective $N_f$ lies much closer to $N_f = 3$. We include these heavy-flavour corrections as described in Ref. \cite{Blumlein2016} with $m_c = 1.59$ GeV and $m_b = 4.78$ GeV.

\subsubsection{Non-perturbative contributions}

Finally, this leaves the treatment of physics of inelastic contributions at and below intermediate momentum scales. There have been three options explored in the literature. The oldest among these (MS) \cite{Marciano2006} takes Eq. (\ref{eq:box_Q2_general}) and defines an interpolation function between the Born amplitude and the DIS regime and requires a matching in $Q^2$ between the Born and DIS regions determined through a fit procedure. The interpolation regime is described using a vector (axial) meson dominance model from large $N_c$ QCD \cite{Marciano2006}, with an effective interaction coming from $\rho, A$ and $\rho'$ mesons. More recent work (DR) \cite{Seng2018, Seng2019b, Gorchtein2018} employed a dispersion relation approach to Eq. (\ref{eq:box_VA_T3}), where $\mathcal{A}_3$ is described by a dispersion integral over a structure function $F_3$, the latter of which is related to experimental (anti)neutrino nucleon scattering through an isospin rotation (cfr. Eqs. (\ref{eq:F3_T3_opt})-(\ref{eq:F_V_DIS})) as discussed above. This allows one to compare model calculations of pion production, Regge physics and resonances in the two-dimensional $(\nu, Q^2)$ space to data. A major finding of the DR results is that the contribution of ``intermediate" scale physics is significantly larger than what was included in MS, and that its influence can be felt even for $Q^2\lesssim 0.1$ GeV$^2$ where the Born term dominates. The idea of separate domains therefore is somewhat flawed, and we must take into account additional hadronic physics not contained in the Born term at low $Q^2$. In response to this, an updated calculation of the original MS results has appeared (CMS) \cite{Czarnecki2019}, which includes additional hadronic effects through a continuation of Eq. (\ref{eq:F_V_DIS}) to lower energy scales. This is done using a number of different methods, including a holomorphic QCD coupling in the infrared for the polarized Bjorken sum rule.

Additional differences in Fermi to Gamow-Teller RC then depend on how (or if) we couple the Born amplitude of Eq. (\ref{eq:box_GT_Born}) to an intermediate regime. In the oldest method (MS), a lower boundary, $Q^2_\mathrm{min}$, is determined by, among others, requiring a smooth continuation such that ${F^V_\mathrm{Born}(Q^2_\mathrm{min})=F^V_\mathrm{INT}(Q^2_\mathrm{min})}$. Because of the larger Born amplitude for the Gamow-Teller contribution, this would imply differences in the fit parameters for $F_\mathrm{INT}$ and $Q^2_\mathrm{min}$, leading to a different interpolation contribution. As shown explicitly by the DR group, however, one of the requirements to constrain $F^V_\mathrm{INT}$ in MS was not valid and additional hadronic physics needs to be included below $Q^2_\mathrm{min}$. A careful treatment using dispersion relations as in Refs. \cite{Seng2018, Seng2019b} would be of great interest, but lies beyond the scope of this work. We follow then an approach similar in spirit to the CMS result, and consider the holomorphic continuation of the GLS and PBj sum rules below $\sim 1$ GeV$^2$. We will additionally go one step further and take into account target mass corrections in the low $Q^2$ domain, and discuss higher-twist corrections.

The QCD sum rules of Eqs. (\ref{eq:GLS_def}) and (\ref{eq:PBjSR_def}) were originally derived in the large $Q^2$ limit following an OPE treatment, far away from the nucleon mass scale at $\sim 1$ GeV$^2$. As one nears this scale, however, several additional contributions arise, known as higher-twist (non-perturbative) and target mass corrections. Both have seen an intense period of research as experimental data became available around and even below the GeV scale \cite{Abe1998, Deur2008}. 

The effect of higher-twist (HT) corrections emerge as a non-perturbative, $1/Q^{2n}$, contribution as one nears the QCD scale. To $\mathcal{O}(1/Q^2)$, contributing matrix elements are typically around the few percent level \cite{Shuryak1982a, Ross1994, Anselmino1995, Stein1995, Blumlein1997, Kataev1998, Kataev2005} at $Q^2=1$ GeV$^2$, depending on the order of the $\alpha_s$ expansion. With regards to the difference between PBj and GLS sum rules (i.e. Fermi and Gamow-Teller RC), however, the situation is not quite as straightforward. In the perturbative domain, it was already mentioned that differences appear only at N$^3$LO due to light-by-light contributions to the GLS sum rule. Initial calculations showed a difference in HT correction terms \cite{Ross1994}, although more recently renormalon results \cite{Kataev2005} show agreement within experimental and theoretical uncertainties. Due to the lack of precise experimental input for the GLS sum rule at low $Q^2$, it is hard to improve upon this point at this time. Explicit chiral perturbation theory calculations might shed light on this issue, which lies however beyond the scope of this work. We will therefore treat its effect only phenomenologically, and encode its influence through a free fit parameter. Additionally, it is not certain that these higher-twist corrections emerge through the isospin rotation unscathed, and we consider their magnitude to come with a 100\% relative uncertainty.

Taking the pQCD expressions described above to even lower momenta ($Q^2 \lesssim 1$ GeV$^2$) becomes increasingly difficult. When taken below $\sim 1$ GeV, the running of $\alpha_s(Q^2)$ using the $\beta$ function explodes and one encounters the Landau pole for which $\alpha_s^{pQCD} \to \infty$ \cite{Deur2016} and which signals the breakdown of pQCD. Several different ways of constructing a holomorphic continuation of $\alpha_s$ into the infrared, using so-called analytical QCD ($\mathcal{A}$QCD), have been explored, and several reviews are available in the literature \cite{Deur2016, Ayala2020}. Because of the large amount of experimental data, we start with a discussion of the PBjSR behaviour, relevant to both axial and vector transitions. We will follow the results of Ref. \cite{Ayala2018a} where different $\mathcal{A}$QCD models were compared to experimental data of the PBjSR after subtraction of the Born contribution (i.e. the $x=1$ contribution in Eq. (\ref{eq:PBjSR_def})). Below a variable threshold, $Q_0^2$, $\mathcal{A}$QCD takes over. Refs. \cite{Ayala2018, Ayala2018a} considered various descriptions of $\Gamma_1^{p-n}$ both below and above $Q_0^2$, and while chiral perturbation theory provides a continuation into the IR, the pQCD+OPE treatment of Eqs. (\ref{eq:T_gamma_W_OPE}) and (\ref{eq:PBjSR_DIS}) was only found to give good agreement with experimental data when using an expression motivated by light-front holography (LFH) \cite{Brodsky2010}. The latter describes the running of the BjSR as follows
\begin{equation}
    1 - \frac{\alpha_{g_1}(Q^2)}{\pi} \stackrel{Q^2 < Q^2_0}{=} 
        1-\exp\left(-\frac{Q^2}{4\kappa^2}\right)
        \label{eq:PBjSR_LFH}
\end{equation}
where $\kappa$ is a fit parameter. While more sophistic models exist in the vicinity of $Q_0^2$, the difference in integrated values are small enough for us to simply use the pQCD+OPE results with the LFH parametrization of Eq. (\ref{eq:PBjSR_LFH}), similar to the CMS approach. Unlike the latter, we leave $\kappa$ to be a free fit parameter.

At intermediate $Q^2$ contributions also appear from discrete resonances. In the case of the GLS sum rule, some complications arise as $F_3^{\nu p (\bar{\nu}p)}$ is an isovector process whereas for our contributions only the isoscalar photonic current contributes. As a consequence, the resonance structure for (anti)neutrino scattering is richer than is the case for us. Luckily, the resonance contribution is very small \cite{Seng2019b}, and we neglect it going forward.

\subsubsection{Target mass corrections}

Turning to target mass corrections, both PBj and GLS sum rules have to be modified when $Q^2$ approaches the nucleon mass scale \cite{Schienbein2008}. Traditionally, this has been performed in two approaches, using either an expansion in $M^2/Q^2$ \cite{Georgi1976}, or a reordering of the OPE coefficients by Nachtmann \cite{Nachtmann1973}. Both approaches are closely related and increase the sum rule predictions for low $Q^2$. Typically, these corrections are removed from experimental results to allow for an extraction of HT contributions and a determination of $\alpha_s$. Here, our purpose is somewhat opposite, since we are interested in the behaviour of Eq. (\ref{eq:box_Q2_general}) over the full $Q^2$ range and all corrections that come with it. In the in low $Q^2$ behaviour, however, an expansion in $M^2/Q^2$ is not very fruitful and we concentrate on the approach by Nachtmann. The latter requires the exchange of the Bjorken-$x$ by
\begin{equation}
    \xi = \frac{2x}{1+\sqrt{1+4x^2M^2/Q^2}}
\end{equation}
which approaches $x$ as $Q^2 \to \infty$. The difference between $x$ and $\xi$ is largest for the elastic contribution ($x=1$), which was already taken into account when discussing the Born term above (Eq. (\ref{eq:F_V_Born})). We use closed expressions for target mass corrections to the $F_3$ and $g_1, g_2$ structure functions as provided in the literature \cite{Wandzura1977, Matsuda1980}, and estimate their effect using simple power law expressions as is performed in Ref. \cite{Kim1998}.

\subsubsection{Numerical results}
In summary we write the total contribution to $F^{A, V}_\mathrm{inel}(Q^2)$ which enters into Eq. (\ref{eq:box_Q2_general}) as
\begin{align}
    F_\mathrm{inel}(Q^2) &= \dfrac{1}{4Q^2} \left\{ \begin{array}{lr}
        1-\dfrac{\alpha_{g_1/F_3}(Q^2)}{\pi} + \dfrac{\mu_4}{Q^2} & \mathrm{(DIS)} \\
        1-\exp\left(-\dfrac{Q^2}{4\kappa^2} \right) & (\mathcal{A}\mathrm{QCD})
    \end{array} \right. \nonumber \\
    &+ F^\mathrm{TMC}_\mathrm{inel}(Q^2),
\end{align}
where $\mu_4$ is the first higher-twist ($\mathcal{O}(1/Q^2)$) contribution.

\begin{figure}
    \centering
    \includegraphics[width=0.48\textwidth]{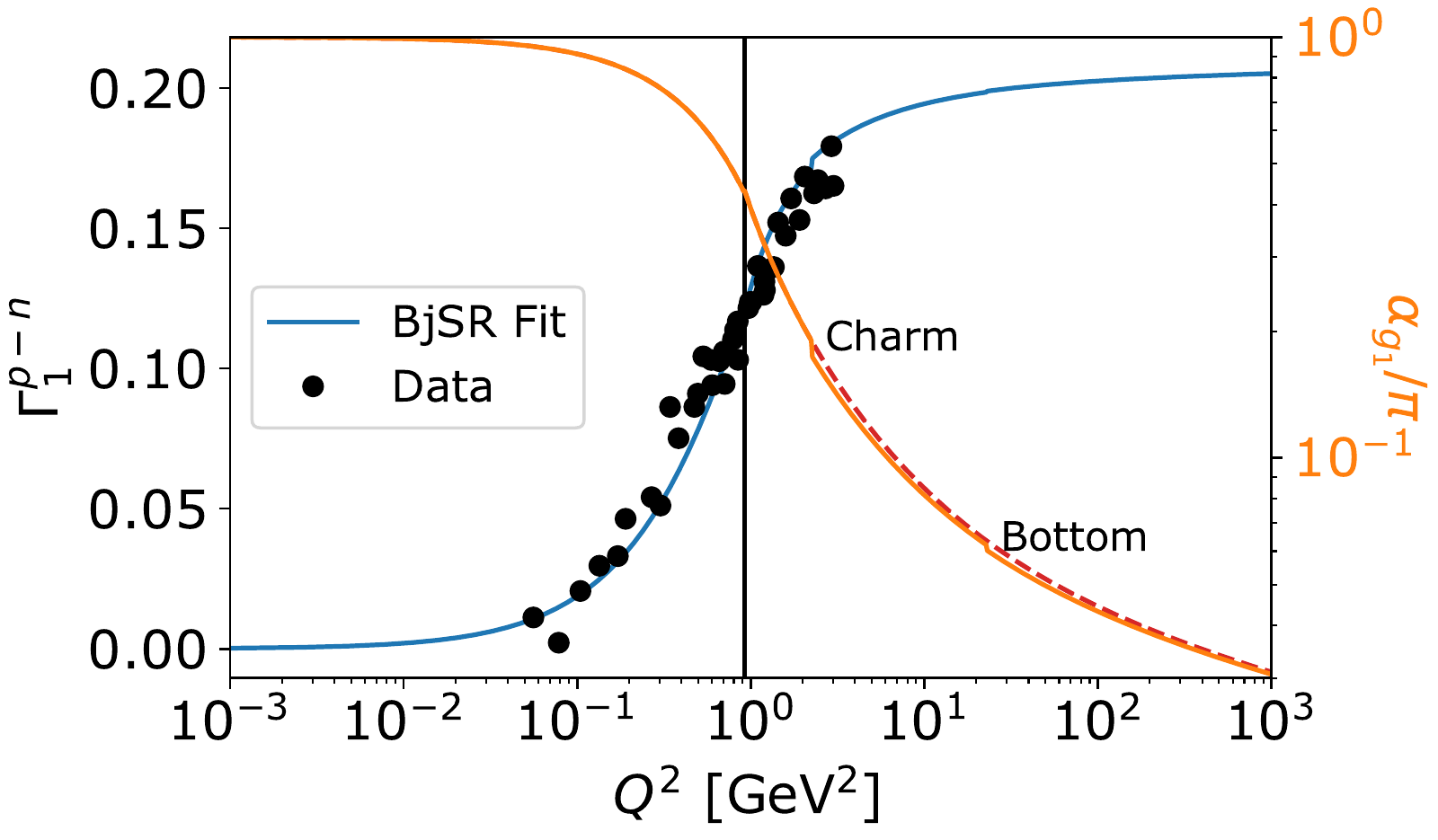}
    \caption{Parametrization of the PBjSR and running coupling $\alpha_{g_1}(Q^2)/\pi$ using the pQCD $\overline{MS}$ parametrization, Eq. (\ref{eq:PBjSR_DIS}), for $Q^2 > Q_0^2 = 0.910$ GeV$^2$ and the LFH result of Eq. (\ref{eq:PBjSR_LFH}) for $Q^2 \leq Q_0^2$, together with experimental data between 0.05 GeV$^2$ and 3 GeV$^2$, adopted from Ref. \cite{Ayala2018a}. The dashed line shows the $N_f=3$ result.}
    \label{fig:PBjSr}
\end{figure}

We use updated input values for the world average of $\alpha_s(M_Z^2) = 0.1179 \pm 0.0010$ \cite{Zyla2020}, a 5 loop $\beta$ function calculation from the \texttt{RunDec} package \cite{Herren2018} and require a smooth transition at $Q_0^2$. For the polarized Bjorken sum rule, our values lie very close to those of Ref. \cite{Ayala2018a} to find $Q_0^2 = 0.910$, $\kappa=0.520 \pm 0.020$ and $\mu_4^\mathrm{Bj} = -0.0221\pm 0.010$, where the latter is the HT contribution of a $1/Q^2$ expansion. This is summarized in Fig. \ref{fig:PBjSr}, where we overlaid the experimental data and show the effect of heavy flavour corrections.

We can perform the same procedure for the GLS sum rule results. Here the available experimental data is much more scarce, however, since these are obtained from (anti)neutrino scattering. A compilation of available data was performed by the CCFR collaboration \cite{Kim1998} for $1.26$ GeV$^2 < Q^2 < 12.59$ GeV$^2$. Since these are still fairly close to the plateau at $Q^2 \to \infty$, however, such a comparison is not a very sensitive probe for the fit parameters as before. Instead, we require continuity in the GLS sum rule and extracted $\alpha_{F_3}(Q^2)$ across $Q_0^2$, where the pQCD results now use the GLS $C_i$ coefficients in Eq. (\ref{eq:alpha_F3}). We find good agreement for $Q^2_0 = 1.05$ GeV$^2$, $\kappa = 0.530 \pm 0.035$ and $\mu_4^\mathrm{GLS}=0.018 \pm 0.025$.

We perform the integration of Eq. (\ref{eq:box_Q2_general}) numerically and find
\begin{align}
    \Box_\mathrm{Bj}^{0} = 0.176(30) \frac{\alpha}{2\pi} &\quad 0 < Q^2 < 0.910\, \mathrm{GeV}^2 \\
    \Box_\mathrm{Bj}^{0} = 2.026(22) \frac{\alpha}{2\pi} &\quad 0.910 \, \mathrm{GeV}^2 < Q^2 < \infty
\end{align}
for the Bjorken sum rule results and 
\begin{align}
    \Box_\mathrm{GLS}^{0} = 0.200(42) \frac{\alpha}{2\pi} &\quad 0 < Q^2 < 1.05\, \mathrm{GeV}^2 \\
    \Box_\mathrm{GLS}^{0} = 2.015(17) \frac{\alpha}{2\pi} &\quad 1.05\,  \mathrm{GeV}^2 < Q^2 < \infty
\end{align}
for the GLS sum rule results, where the superscript ``0'' denotes the omission of TMC. The uncertainties arise from the change in fit parameters and a 100\% uncertainty on the higher twist contributions. The contribution of heavy-flavour corrections is $\mathcal{O}(10^{-5})$, but we include it for completeness.

Finally then, the target mass corrections are implemented as described above, and change the box contribution with
\begin{align}
    \Box_\mathrm{Bj}^\mathrm{TMC} = 0.089(45) \frac{\alpha}{2\pi} &\quad 0 < Q^2 < 0.910\, \mathrm{GeV}^2 \\
    \Box_\mathrm{Bj}^\mathrm{TMC} = 0.022(11) \frac{\alpha}{2\pi} &\quad 0.910 \, \mathrm{GeV}^2 < Q^2 < \infty
\end{align}
for the Bjorken sum rule and 
\begin{align}
    \Box_\mathrm{GLS}^\mathrm{TMC} = 0.092(46) \frac{\alpha}{2\pi} &\quad 0 < Q^2 < 1.05\, \mathrm{GeV}^2 \\
    \Box_\mathrm{GLS}^\mathrm{TMC} = 0.017(9) \frac{\alpha}{2\pi} &\quad 1.05\,  \mathrm{GeV}^2 < Q^2 < \infty
\end{align}
for the GLS sum rule results. Since the behaviour of the GLS and PBj sum rules is identical to leading order, the target mass corrections are common within uncertainties and increase both results almost equally. We have conservatively estimated our uncertainties at 50\% of the magnitude of the effect. Note that in this case, the shift corresponds to more than 1 sigma when compared to the CMS results, who took the uncertainty on the $Q^2 < 1.1$ GeV$^2$ region to be a blank 20\%.

\begin{figure}[ht]
    \centering
    \includegraphics[width=0.48\textwidth]{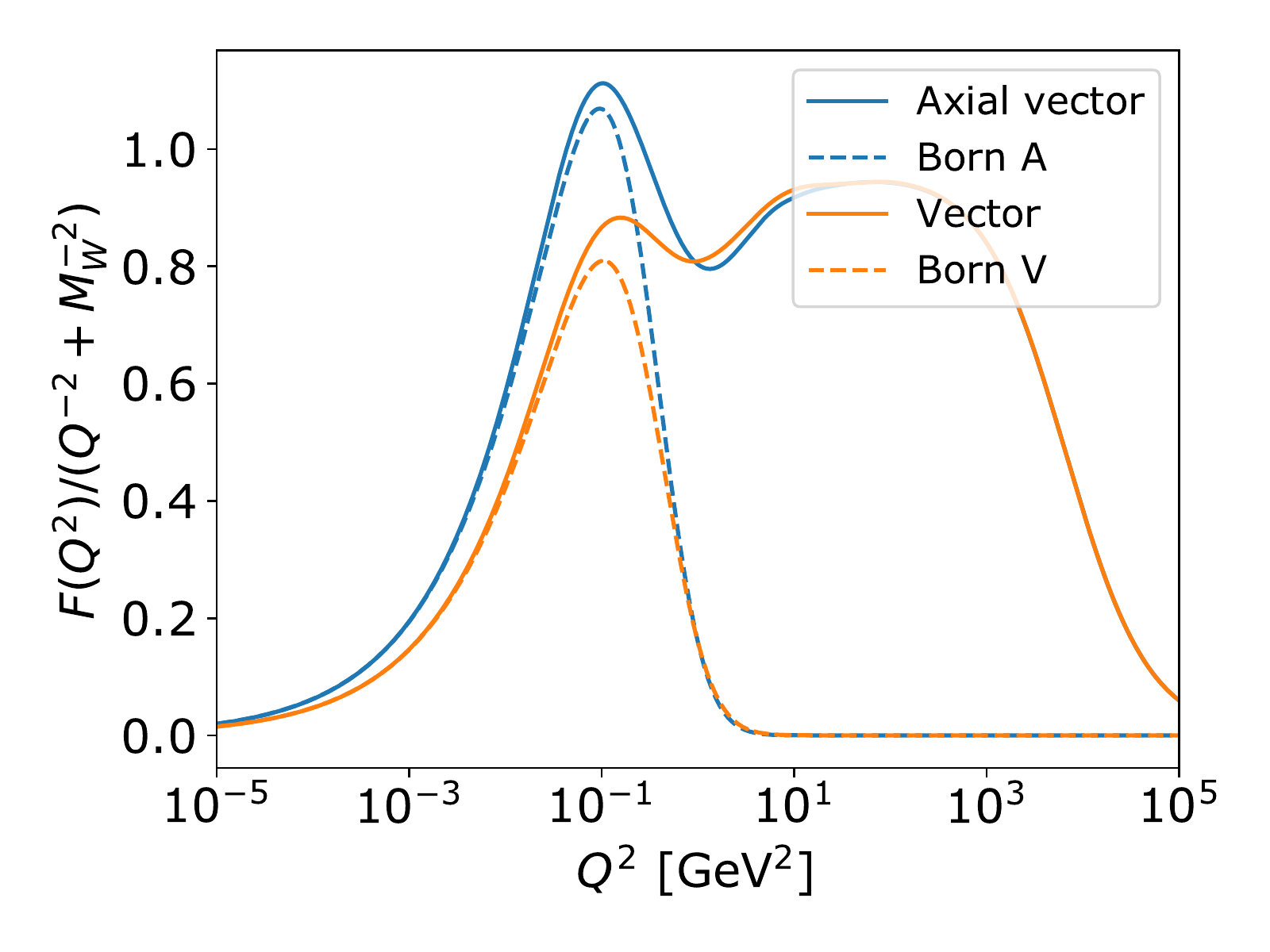}
    \caption{Summary of the results for Vector and Axial Vector transitions including target mass corrections, calculated as in Ref. \cite{Seng2019b}. Dashed lines show the contribution of the Born amplitude.}
    \label{fig:Mellin_Seng}
\end{figure}

In our discussion above we have alluded to the possibility of using either GLS or PBj sum rule results for the vector transition, with the argument relying either on isospin or chiral symmetry, respectively. In Ref. \cite{Czarnecki2019} one takes the PBjSR results also below $1.1$ GeV$^2$, i.e. in the regime where chiral symmetry is expected to broken. In the DR work \cite{Seng2018, Seng2019b}, one uses isospin symmetry to relate it to the $F_3^{\nu p(\bar{\nu} p)}$ structure function. As also shown in the Appendix, this correspondence is not completely model-independent since the $\gamma W$ contribution is of the isoscalar type, whereas (anti)neutrino scattering is fully isovector. Both in the elastic channel and for intermediate (Regge \cite{Collins1977}) momentum scales, this correspondence can be clearly established. In the DIS regime, the small difference between GLS and PBj sum rules provides additional credence to this hypothesis, and the authors of Ref. \cite{Seng2019b} conclude this translation can be made up to isospin breaking ($\sim$ few percent) corrections. We follow the same philosophy here, but use the $\mathcal{A}$QCD continuation of the GLS sum rule to capture the low $Q^2$ behaviour coupled with the PBjSR DIS regime. Additional details are provided in the Appendix. 

Our results are summarized in Fig. \ref{fig:Mellin_Seng}, shown in a way similar to Ref. \cite{Seng2019b}. We see that the holomorphic results for a vector transition resemble the DR results much closer than the original MS results, shown in Fig. 7 of Ref. \cite{Seng2019b}. The increase in the Born amplitude for the axial transition is clearly visible, even though the difference due to intermediate scale physics from the difference in GLS and Bj sum rules is not statistically significant. This is not surprising, given that they approach each other in the chiral limit, and the lack of high precision data for the GLS sum rule allows for large variations. Target mass corrections further lift the response at low energies, predominantly around $Q^2 \lesssim 0.1$ GeV$^2$. We note that chiral breaking effects will likely play a role at low $Q^2$ for a difference in $\Delta_R^{V, A}$, which is a topic of further study.

\section{Effective couplings}
\label{sec:summary_order_alpha}
\subsection{Nucleons}
We have identified three sources of $\mathcal{O}(\alpha)$ radiative corrections that are \textit{a priori} different for Fermi to Gamow-Teller transitions. Two of these originated from the non-zero divergence of the axial current, Eq. (\ref{eq:D_a_3p}) and (\ref{eq:D_2pt}). In both cases the UV contribution vanished, which can be intuitively understood from the partially conserved axial current hypothesis. Somewhat more surprising is that also the Born contribution vanishes, either through a cancellation between isoscalar and isovector parts (Eq. (\ref{eq:D_gamma_cancellation_scalar_vector})) or crossing symmetry for the isovector contribution (\ref{eq:D_2p_Born}). The only remaining $\mathcal{O}(\alpha)$ difference was found to originate in the vector induced part of the $\gamma W$ box. Specifically, we found an increase in the Born contribution for Gamow-Teller transitions due to the influence of weak magnetism in the weak nucleon vertex, Eq. (\ref{eq:box_GT_Born}). We have treated all other non-elastic contributions based on the polarized Bjorken and Gross-Llewellyn Smith sum rules, using pQCD for $Q^2 \gtrsim 1$ GeV$^2$ and a holomorphic continuation towards the infrared using light front holography results, constrained by experimental data and continuity requirements. We have supplemented these results using highest-twist and target mass corrections, with changes to numerically integrated values predominantly arising from the latter. Since the running of the two sum rules coincide in the chiral limit, it is unsurprising that their difference is small, and not statistically significant. 

For the total inner RC we use the expressions obtained from summing large logs using renormalization groups \cite{Hardy2015, Czarnecki2019}
\begin{equation}
    \Delta_R = 0.01671 + 1.022 A_{NP} + 1.065A_P
\end{equation}
where the first term corresponds to all common, model-independent logarithmic factors of Eq. (\ref{eq:inner_vector_blown_up}) and $A_{(N)P}$ are (non-)perturbative contributions discussed in the previous section. Summing everything together we have
\begin{align}
    \Delta_R^V &= 0.02473(27) \label{eq:Delta_R_V_final} \\
    \Delta_R^A &= 0.02532(22) \label{eq:Delta_R_A_final}
\end{align}
We note that $\Delta_R^V$ agrees nicely with the dispersion relation results of Refs. \cite{Seng2018, Seng2019b}. It is somewhat larger than the new results of Czarnecki, Marciano and Sirlin \cite{Czarnecki2019}, which can be traced back to two different effects. The first is because we argue that the Born contribution should be integrated up to $Q^2 \to \infty$ rather than the cutoff energy at which pQCD contributions arise, similar to the dispersion relation results and the treatment of the QCD sum rules upon which their analysis was based. Second, the contributions due to target mass corrections are substantial mainly in the low $Q^2$ domain and increase results non-trivially. By including these corrections, the dispersion results are very similar in spirit to the ones we have presented here. Both rest on the argument that in the isospin limit, we can identify expressions will well-studied QCD sum rules. While the dispersion results go to great lengths to motivate their physics input over the entire domain, the analytical continuation presented here must be consistent with a subset of the same data that Ref. \cite{Seng2019b} is comparing to. It is therefore hardly surprising that in the end our results agree.

Our uncertainty is larger than the DR results, but smaller than those of CMS. Taking a closer look at the latter, the predominant source of uncertainty arises almost equally from the blanket 5\% and 10\% relative uncertainty on the DIS and Born contributions, respectively. In the DR result, on the other hand, no uncertainty is provided for the DIS contribution and the uncertainty on the Born amplitude is derived from data. Here we decided to take an intermediate approach, with the uncertainty on the Born contribution in accordance with the DR work but an uncertainty on the DIS regime due to fit uncertainties and a 100\% relative uncertainty on higher-twist corrections.

The difference in inner radiative corrections between vector and axial vector is now found to be
\begin{equation}
    \Delta_R^A-\Delta_R^V = 0.52(5) \frac{\alpha}{2\pi} = 0.60(5) \times 10^{-3}.
    \label{eq:change_RC_neutron}
\end{equation}
where the uncertainty originates from the form factors in the Born contribution and the ambiguity in GLS non-elastic results taken in quadrature. Since the target mass corrections are the same within uncertainties and strongly correlated we do not take its additional error into account. The difference is then driven almost exclusively by the elastic response, in particular that of that of the weak magnetism contribution.

This also allows one to, for the first time, extract the underlying $g_A$ from experimental measurements which is to be used in neutral current processes and used in comparison with lattice QCD. Using the most precise individual measurement \cite{Markisch2019}, $g_A^\mathrm{eff} = 1.27641(56)$, we find 
\begin{align}
    g_A^{0} &\equiv \dfrac{g_A^\mathrm{eff}}{1+(\Delta_R^A-\Delta_R^V)/2} \\
    &=1.27603(56),
\end{align}
or a $0.7\sigma$ shift with respect to the traditionally quoted value.

\subsection{Nuclear effects}
\label{sec:nuclear_effects}
Up to now, we have treated only the case where the initial and final nucleon in the diagrams of Fig. \ref{fig:feynman_order_alpha} are the same nucleon. In a nucleus, however, this need not be the case. As a consequence, an additional term shows up which depends on nuclear structure \cite{Hardy2015}
\begin{equation}
    1+\Delta_R \to (1+\Delta_R)(1-\delta_C+\delta_{NS})
\end{equation}
where $\delta_C$ are so-called isospin breaking corrections and $\delta_{NS}$ is the effect of multiple nucleons in the $\gamma W$ box diagram. For the case of superallowed $0^+\to 0^+$ Fermi transitions explicit calculations have been performed, taking into account two different nucleons in initial and final state \cite{Towner1992}. There it was found that in general the corrections depend on
\begin{equation}
    \delta_{NS}^F \sim \frac{\langle p_N \rangle}{M} = \frac{v_N}{c},
\end{equation}
where $\langle p_N \rangle$ is the average nucleon momentum and $v_N$ the corresponding velocity. This can be intuitively understood since the Fermi transition receives contribution from the axial vector part of $T^{\mu\nu}$. Because of the contraction with the asymmetric tensor at least one index must be spacelike, so that the amplitude for nucleons depends on $v_N/c$. The same argument applies for a Gamow-Teller transition, so that \textit{a priori} the contributions are expected to be of similar size.

Another way of treating nuclear structure information has traditionally been achieved via the decomposition of the weak vertex, $W^\mu$ in Eq. (\ref{eq:current_decomp_nucleon}), into model-independent form factors in one of two ways. The first is to perform a spherical tensor decomposition in the Breit frame, where the timelike and spacelike currents can separately be expanded using (vector) spherical tensors \cite{Stech1964, Schulke1964, Behrens1971, Behrens1982}. The other consists of a manifest Lorentz invariant decomposition, which is practical mainly for allowed decays due to the limited amount of terms \cite{Holstein1974}. For the purpose of the discussion here, we use the latter for its clarity, even though the results obtained using the former will be identical (up to $\mathcal{O}(q/M)$). All nuclear structure information is then encoded into form factors. In this case we can write \cite{Holstein1974}
\begin{align}
    V_\mu(q) &= \frac{1}{2M}(aP_\mu + eq_\mu)\delta_{JJ'}\delta_{MM'} + i \frac{b}{2M}\epsilon_{0i\mu k}q^i\mathcal{C}_1^k \nonumber \\
    &+\frac{\mathcal{C}_2^k}{2M}\biggl[\text{higher order} \biggr] \label{eq:V_HS} \\
    A_\mu(q) &= \frac{\mathcal{C}_1^k}{4M}\epsilon_{ijk}\epsilon_{ij\mu \nu} \biggl[cP^\nu - dq^\nu + \ldots \biggr] \nonumber \\
    & + \frac{\mathcal{C}_{2,3}^k}{(2M)^2} \biggl[ \text{higher order} \biggr]
    \label{eq:A_HS}
\end{align}
where $C_i^k$ is a Clebsch-Gordan coefficient, $P = p_i + p_f$ and all form factors are a function of $q^2$. Typically, the form factors are expanded using a power series in $q^2$, or assumed to be of a dipole shape. This then usually corresponds to including only the Born contribution and discussed in the previous section. This serves as the replacement of Eq. (\ref{eq:current_decomp_nucleon}). In the case of the neutron the correspondence can be read off directly from comparing the latter and Eqs. (\ref{eq:V_HS}) and (\ref{eq:A_HS}), where the higher-order terms are zero. The calculation then proceeds analogously as for the neutron, and assuming a dipole shape for the form factors one finds \cite{Holstein1979}
\begin{align}
    \Delta_R^A-\Delta_R^V \sim \frac{4}{5}\frac{\alpha Z}{MR}\frac{b}{Ac},
    \label{eq:change_RC_wm_nuclear}
\end{align}
where $R=\sqrt{5/3}\langle r^2\rangle^{1/2}$ is the nuclear radius, $Z$ is its atomic number and $b (c)$ is the so-called weak magnetism (Gamow-Teller) form factor. We can understand the appearance of the factor $\alpha Z$ rather than $\alpha$ as follows. While in theory every nucleon inside a nucleus can undergo decay, because of their occupancy in specific orbitals and relative position with respect to the Fermi energy, only those closest to the latter do at a reasonable rate. When two different nucleons are involved, however, every nucleon which interacts with the outgoing $\beta$ particle through exchange a photon can do so equally, with the other nucleon near the Fermi energy interacting with the $W$ boson. Besides this simplified picture additional effects show up. This is in part because of the presence of discrete levels at the MeV rather than GeV scale and a significant quasi-elastic response \cite{Seng2019b, Gorchtein2018}. While these effects can be expected to be of similar magnitude, a more detailed treatment lies beyond the scope of this work.

\section{The lattice and right-handed currents}
\label{sec:gA_BSM}
Traditionally, one defines $g_A$ as in Eq. (\ref{eq:gA_def_QCD}), i.e. containing any difference in vector to axial RC and potential BSM signals. Because of the rapid progress in the field of lattice QCD, an accurate first principles calculation of $g_A^\mathrm{QCD}$ has been demonstrated to the percent level \cite{Chang2018, Gupta2018}, although it is currently unclear how some systematic effects influence the final accuracy \cite{Aoki2020}. Nevertheless, a comparison between experimentally obtained values for $\lambda \equiv g_A/g_V$ and calculations for $g_A^\mathrm{QCD}$ allow one to disentangle potential BSM signatures in a clean system. Assuming new charged current physics to appear only at high scales, $\Lambda^2_\mathrm{BSM} \gg M_W^2$, we can treat the problem using an effective field theory \cite{Bhattacharya2012, Cirigliano2013, Cirigliano2013b, Naviliat-Cuncic2013, Gonzalez-Alonso2018}
\begin{equation}
    \lambda_{EFT} = \lambda_{SM}(1-2\,\mathrm{Re}\,[\epsilon_R])
    \label{eq:lambda_EFT}
\end{equation}
where $\epsilon_R$ is a BSM right-handed coupling constant assuming new UV physics, interpreted in the Standard Model EFT. Within the context of BSM searches in the charged current sector, the particular form of Eq. (\ref{eq:lambda_EFT}) is pleasing because of its simplicity and sensitivity enhancement. On the other hand, a difference in radiative corrections between vector and axial vector transitions mimics exotic right-handed currents, so that a failure to take it into account would incorrectly lead to a non-zero BSM signal when the precision reaches the expected offset. Using the results of Eq. (\ref{eq:change_RC_neutron}), we find
\begin{align}
    \lambda_{SM} &= \frac{g_A^{QCD}}{g_V} \left[1 + \frac{1}{2}(\Delta_R^A-\Delta_R^V)\right] \\
    &= \frac{g_A^{QCD}}{g_V}[1+0.30(3)\times 10^{-3}].
    \label{eq:lambda_renormalization}
\end{align}

As a consequence, experimental results extract $\lambda_{EFT}$ \cite{Markisch2019, Brown2018, Gonzalez-Alonso2018}, which is then assumed to be equal to $g_A^\mathrm{QCD}$ after setting $g_V$ to unity \cite{Ademollo1964}. We find that the difference is smaller than 0.1$\%$.

\begin{figure}[!ht]
    \centering
    \includegraphics[width=0.48\textwidth]{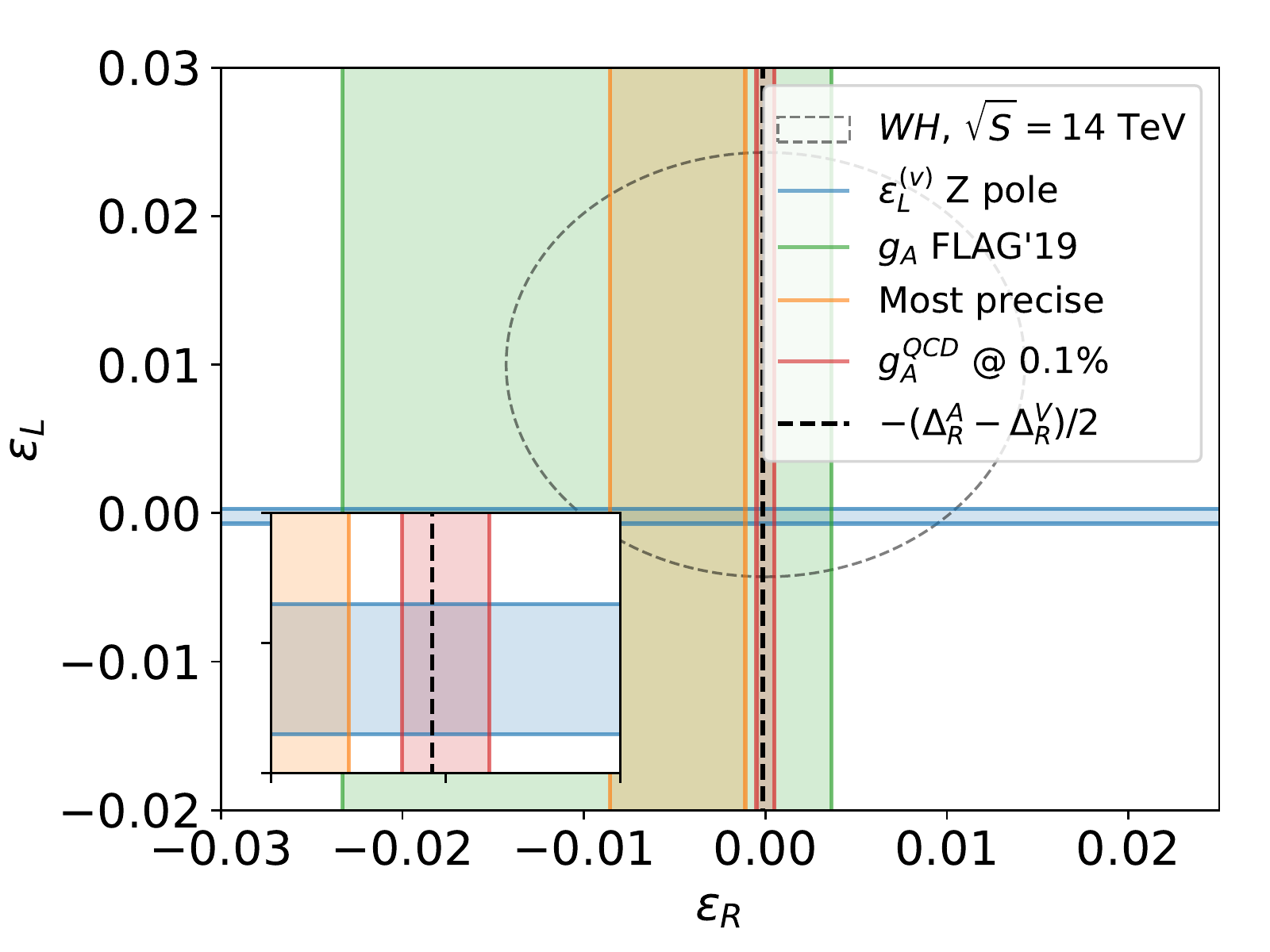}
    \caption{Current limits (68\% C.L.) on left and right-handed couplings interpreted in the SMEFT, showing $Z$-pole (blue) \cite{Falkowski2017, Efrati2015}, LHC (black) \cite{TheATLAScollaboration2016}, LQCD results from FLAG'19 \cite{Aoki2020} and Ref. \cite{Walker-Loud2020}. In red we show anticipated limits when $g_A$ reaches 0.1\% on the lattice. The black vertical line represents the effects of Eq. (\ref{eq:lambda_renormalization}) as a false BSM signal.}
    \label{fig:epsilonR_limits}
\end{figure}

Currently, there are a number of results available for a LQCD determination of $g_A$. We compare here two different results: The FLAG 2019 summary \cite{Aoki2020}, which finds $g_A^{FLAG} = 1.251(33)$ and the most precise (MP) individual determination published this year, $g_A^{MP} = 1.2642(93)$ \cite{Walker-Loud2020}. The calculated shift in $g_A$ from Eq. (\ref{eq:lambda_renormalization}) corresponds to about 1/3rd of the MP result. The anticipated shift of Eq. (\ref{eq:lambda_renormalization}) and the possibility of detecting right-handed currents through $\lambda$ has prompted interest in pushing for a more precise calculation in the near future \cite{Walker-Loud2020}. Figure \ref{fig:epsilonR_limits} shows the current and anticipated limits using $g_A$ from the lattice with the recent PDG average for $\lambda = 1.2756(13)$ \cite{Zyla2020}.

The correction corresponds to a 0.02\% shift in $\epsilon_R$, which leaves the current limits unchanged due to the large uncertainty of lattice results for $g_A^\mathrm{QCD}$. As mentioned above, however, there is significant interest in improving the precision of the latter \cite{Walker-Loud2020}. After correcting for Eq. (\ref{eq:lambda_renormalization}), equality between experimental and lattice values for $g_A$ will then put the most stringent direct limits on right-handed currents\footnote{We have omitted here the combination of CKM unitarity ($\Delta_{CKM} \propto \epsilon_L + \epsilon_R$) and the pion decay ($\delta \Gamma_{\pi \to \mu 2} \propto \epsilon_L - \epsilon_R$) due to the degeneracy with pseudoscalar, scalar, and tensor interactions \cite{Alioli2017, Gonzalez-Alonso2018}.}. 

We note again that although the relative difference between $\Delta_R^A$ and $\Delta_R^V$ is relatively small, the Born contribution to the bare $\gamma W$ integral is increased by almost $70\%$ for the axial vector renormalization, and should be accessible via LQCD calculations with an explicit photon. 

\section{Consistency issues in traditional \texorpdfstring{$\beta$}{beta} decay theory input}
\label{sec:traditional_beta_decay}
Upon closer inspection, some of the results obtained in traditional $\beta$ decay formalisms \cite{Holstein1974, Behrens1982, Hayen2018} have the same origin as some of the radiative corrections discussed above, although the connection is not immediately clear when comparing final expressions. Because the neutron calculations do not have take into account any nuclear response, calculations can be performed in a more straightforward manner and historically results have been published using several different formalisms. On the nuclear theory side, the connection with radiative corrections is typically not as obvious in the formalisms that are commonly used, and the main QED effect that is taken into account is the Coulomb interaction. The latter can be understood as part of the low $k \ll M_W$ contribution of the $\gamma W$ box diagram of Sec. \ref{sec:gamma_W_box}. While this is obvious for the leading Coulomb term ($\sim \alpha Z/ \beta$ with $\beta = v/c$ the velocity), additional higher-order terms sneak in. Some of these cancel in the full $\mathcal{O}(\alpha)$ calculation as we have shown above, while they survive in the traditional $\beta$ decay results. Further, because some of these additional terms are included in some elements of the commonly used theory input and not in others for, e.g., correlation measurements in nuclear mirror systems, double counting occurs when putting all results together for, e.g., a $V_{ud}$ extraction.

\subsection{Missing cancellation}
\label{sec:cancellation}
In the traditional $\beta$ decay calculations of the second half of the last century \cite{Holstein1974, Behrens1982}, a particular focus was placed on a rigorous classification of the nuclear current while taking into account the Coulomb interaction between initial and final state as the dominant QED correction. In the Standard Model this is to be understood to first order in $\alpha Z$ as the Born amplitude of the $\gamma W$ box, using only the electric monopole term. Taking Eq. (\ref{eq:gamma_W_box_init}) and using the Born amplitude of Eq. (\ref{eq:T_munu_Born}), to first order in $\alpha Z$ the matrix element can be written as follows
\begin{align}
    &4 \pi \alpha G_FV_{ud} \int \frac{d^4k}{(2\pi)^4} \frac{\bar{e}(2l^\mu - \gamma^\mu\slashed{k})\gamma^\nu (1-\gamma^5)\nu}{k^2[k^2 - 2l\cdot k]}\nonumber \\
    & \times \bar{u}\left[Z F_1^f(k^2)\frac{2p_{f,\mu} + \gamma_\mu\slashed{k}}{k^2+2p_f\cdot k + i\epsilon} W_\nu(p_i+k, p_f) \right. \nonumber \\
    &\left.+(Z-1)F_1^i(k^2) W_\nu(p_i, p_f-k)\frac{2p_{i,\mu} - \gamma_\mu \slashed{k}}{k^2-2p_i\cdot k + i\epsilon} \right]u.
\end{align}
Neglecting the difference between $Z$ and $Z-1$ and assuming the normalized charge form factors, $F_1^{i,f}(k^2)$ to be the same (analogous to taking only the isoscalar moment as we have done above), using that $p_f \approx p_i = (M, \bm{0})$ in the center of mass frame and neglecting $k \ll M$ due to the suppression of the form factors for high $k^2$, one arrives at
\begin{align}
    &-i4 \pi \alpha Z G_FV_{ud} \int \frac{d^4k}{(2\pi)^4} \frac{\bar{e}(2p^0 - \gamma^0\slashed{k})\gamma^\nu (1-\gamma^5)\nu}{k^2[k^2 - 2p\cdot k]}\nonumber \\
    & \times 2MF_1 \bar{u}\left[\frac{W_\nu(p+k, p)}{k^2-2p\cdot k + i\epsilon}+\frac{W_\nu(p, p-k)}{k^2-2p\cdot k + i\epsilon} \right]u
    \label{eq:matrix_element_traditional_0}
\end{align}
Using Eq. (\ref{eq:propagator_res_trick}) to reduce the hadronic propagators and recognizing now the definition of the Coulomb potential to order $\alpha Z$ \cite{Holstein1979b}
\begin{equation}
    V_C(\bm{r}) = 8\pi \alpha Z\int \frac{d^3k}{(2\pi)^4}\frac{1}{\bm{k}^2}e^{i\bm{k}\cdot \bm{r}}F_1(k^2)
\end{equation}
the electron wave function to order $\alpha Z$ is then
\begin{equation}
    \bar{\phi}_e(\bm{r}, \bm{p}) = \bar{u} e^{-i\bm{p}\cdot \bm{r}} - i \int d^4z\, \bar{u}e^{ipz}\gamma_0 V_C(\bm{z}) S_F(z-r)
\end{equation}
with $S_F$ the fermion propagator. One then generalizes the resulting form to take $\bar{\phi}_e$ as the solution to the Dirac equation in the central Coulomb potential of the daughter to all orders in $\alpha Z$. Finally, we obtain the traditional Coulomb-corrected $\beta$ decay amplitude amplitude as first written down by Stech and Sch\"ulke \cite{Stech1964, Holstein1979b},
\begin{align}
    \mathcal{M}_{fi} &= \int \mathrm{d}^3 r\, \bar{\phi}_e(\bm{r}, \bm{p}_e)\gamma^\mu(1-\gamma^5)v(\bm{p}_{\bar{\nu}}) \nonumber \\
    & \times \int \frac{\mathrm{d}^3s}{(2\pi)^3}e^{i\bm{s}\cdot \bm{r}}\frac{1}{2}[\langle f(\bm{p}_f+\bm{p}_e-\bm{s})| V_\mu + A_\mu | i(\bm{p}_i) \rangle \nonumber \\
    & + \langle f(\bm{p}_f) | V_\mu + A_\mu | i(\bm{p}_i-\bm{p}_e+\bm{s}) \rangle ].
    \label{eq:generalized_matrix_element_SS}
\end{align}
The vector and axial vector currents can then be replaced by, e.g., Eqs. (\ref{eq:V_HS}) and (\ref{eq:A_HS}) or a (vector) spherical harmonics expansion as is done in the work of Behrens and B\"uhring \cite{Behrens1982}. Upon inspection, it is clear that $\bm{s} = \bm{p}_e-\bm{k} \approx -\bm{k}$ for large loop momenta. The calculation then proceeds through a similar expansion of the lepton current which defines the basic matrix element. While this in itself is not a problem, based on our discussion of the Born term in Sec. \ref{sec:parity_odd} is it clear that for $p_e \ll \bm{k} \ll M$ terms of $\mathcal{O}(\alpha Z/MR)$ show up, see Eq. (\ref{eq:change_RC_wm_nuclear}). This had been noted before \cite{Bottino1974, Holstein1974b} and is included by default in the Behrens-B\"uhring formalism even though there was no explicit publication of the latter. In particular, it was observed that a renormalization of sorts happens to the different form factors, such as for the Gamow-Teller form factor \cite{Holstein1979, Hayen2018}
\begin{equation}
    c \to c \pm \frac{2}{5}\frac{\alpha Z}{MR} \frac{\pm 2b + d}{Ac},
    \label{eq:dumb_renormalization}
\end{equation}
with $b, c$ and $d$ the weak magnetism, Gamow-Teller and induced tensor form factors in the Holstein notation as in Eqs. (\ref{eq:V_HS}) and (\ref{eq:A_HS}). What is of special importance, however, is that the origin of the $b$ and $d$ terms differ, as they originate from different terms of the reduction of the product of three gamma matrices in Eq. (\ref{eq:matrix_element_traditional_0}) when using Eq. (\ref{eq:triple_gamma_reduction}). We find that the $d$ term arises from the piece equivalent to $\slashed{k}T^0{}_0$ in Eq. (\ref{eq:gamma_W_box_WTI}), whereas the weak magnetism contribution arises from the parity-odd amplitude, $\epsilon_{\mu\nu\alpha\beta} k^\alpha L^\beta T^{\mu\nu}$, as we have seen above. In the full calculation, however, the former cancels completely with the low-energy part of the vertex correction, see the discussion at Eq. (\ref{eq:fermion_propagator_trick}) and the appendix. As a consequence, the $(\alpha Z/ MRc) d$ term should not be present in a consistent $\mathcal{O}(\alpha)$ calculation,
\begin{equation}
    \frac{2}{5}\frac{\alpha Z}{MR}\frac{d}{Ac} \to 0,
\end{equation}
and care must be taken when combining $\mathcal{O}(\alpha)$ radiative corrections calculations with classical calculations of the $\beta$ decay rate such as those listed in Refs. \cite{Holstein1974, Hayen2018}. For Fermi transitions this is not a problem, as even in the ``naive'' calculation of Eq. (\ref{eq:generalized_matrix_element_SS}) the total contribution vanishes.

\subsection{\texorpdfstring{$|V_{ud}|$}{Vud} Double counting in \texorpdfstring{$T=1/2$}{T=1/2} mirror decays}
\label{sec:mirror_Vud}

The second issue pertains to the evaluation of $V_{ud}$ from mirror decays, i.e. $\beta$ transitions within an isospin $T=1/2$ doublet. The master equation relating the lifetime, phase space and matrix elements can be obtained by making the substitution $3\lambda^2 \to \rho^2$ in Eq. (\ref{eq:general_tiple_relation}) and inserting the Fermi matrix element, $M_F$,
\begin{align}
    t_{1/2}f_V\left[1+\frac{f_A}{f_V}\rho^2\right] = \frac{2\pi^3\hbar \ln 2}{M_F^2V_{ud}^2G_F^2g_V^2(m_ec)^5}\frac{1}{1+RC}
    \label{eq:triple_relation_mirror}
\end{align}
where we have inserted the half-life rather than lifetime and 
\begin{equation}
    \rho = \left\{
        \begin{array}{ll}
        \dfrac{c(q^2)}{a(q^2)} & \text{Holstein \cite{Holstein1974}} \\
        \dfrac{{}^AF_{101}(q^2)}{{}^VF_{000}(q^2)} & \text{Behrens-B\"uhring \cite{Behrens1982}}
    \end{array} \right.
\end{equation}
is the ratio of Gamow-Teller and Fermi form factors in the two most popular formalisms\footnote{Depending on the formalism, the sign of $\rho$ can change. Since we are concerned here only with $\rho^2$ we refer the reader to, e.g., \cite{Hayen2020a} for more detail.}. Because its decay occurs within an isospin doublet, the Fermi matrix element is completely determined thanks to the conservation of the weak vector current and one finds $M_F^0 = 1$, where the superscript denotes the assumption of isospin symmetry. In this sense, it can be thought of as the nuclear equivalent of the neutron which brings with it a number of distinct advantages. As with the neutron, $\rho$ can be determined experimentally through $\beta(-\nu)$ correlation measurements, with some isotopes gaining significant enhancements due to near-cancellations \cite{Hayen2020a}. In summary, we can define the so-called corrected $ft$ value common to all mirror decays (i.e. all the nucleus-independent factors in the rhs of Eq. (\ref{eq:triple_relation_mirror})), $\mathcal{F}t_0$, which is defined as \cite{Naviliat-Cuncic2009}
\begin{align}
    \mathcal{F}t_0 &= g_V^2f_Vt(1+\delta_R^\prime)(1+\delta^V_{NS}-\delta_C^V) [1+(f_A/f_V)\rho^2] \nonumber \\
    &\equiv \mathcal{F}t[1+(f_A/f_V)\rho^2],
    \label{eq:Ft_0_mirror}
\end{align}
where $\delta_i$ are outer radiative ($R$), nuclear structure ($NS$) and isospin-breaking ($C$) corrections \cite{Severijns2008}, following $|M_F|^2 = |M_F^0|^2(1+\delta_C^V)=1+\delta_C^V$ \cite{Hardy2009}. Then, if theory input is provided for the so-called phase space factors $f_{A,V}$, one can extract a complementary determination of $V_{ud}$, the up-down quark mixing matrix element \cite{Severijns2008, Naviliat-Cuncic2009} from the relation
\begin{equation}
    V_{ud}^2 = \frac{K}{\overline{\mathcal{F}t}_0G_F^2(1+\Delta_R^V)}
    \label{eq:Vud_mirrors_general}
\end{equation}
where $K/(\hbar c)^6 = 2\pi^3\, \ln{2\hbar}/(m_ec)^5 = 8120.278(4) \times 10^{-10}\,$GeV$^{-4}\,$s, $G_F/(\hbar c)^3 = 1.1663787(6) \times 10^{-5}\,$GeV$^{-2}$ \cite{Tishchenko2013} and $\Delta_R^V = 2.467(22)\%$ the inner radiative correction obtained from dispersion relations \cite{Seng2018, Seng2019b} or our own result in Eq. (\ref{eq:Delta_R_V_final}).

The problem now is the following: the quantities $f_{A, V}$ are calculated as the integral of the $\beta$ spectrum shape for vector and axial vector transitions in the Behrens-B\"uhring formalism \cite{Hardy2015, Towner2015, Hayen2018}, whereas experimental analyses typically use expressions based on that of Holstein \cite{Holstein1974} or older resources to extract $\rho$. As we have seen in the previous section, parts of the Gamow-Teller-specific RC by default leak into the formalism in the former, whereas these have to be added \textit{post-hoc} in the latter \cite{Holstein1979}, and which are not included in experimental analyses and compilations of formulae. As a consequence, the analysis of experimental data returns $\rho_{SM}$ - which includes the renormalization analogous to Eq. (\ref{eq:lambda_renormalization}) - so that when it is combined into Eq. (\ref{eq:Ft_0_mirror}) double-counting occurs\footnote{It is somewhat fortuitous that the effect is smaller than it could have been since in Eq. (\ref{eq:dumb_renormalization}) $d=0$ for decays within isospin multiplets.}. 

We recalculate the standard $f_A/f_V$ values \cite{Severijns2008, Naviliat-Cuncic2009, Towner2015} by subtracting the $\alpha Z/(MRc)b$ contributions to the result. Table \ref{tab:new_Vud_mirrors} lists updated $f_A/f_V$ and $\mathcal{F}t_0$ values for the isotopes for which all experimental information is available to allow extraction of $V_{ud}$: $^{19}$Ne, $^{21}$Na, $^{29}$P, $^{35}$Ar and $^{37}$K.

\begin{table}[!ht]
    \centering
    \begin{ruledtabular}
    \begin{tabular}{r|llll}
    & $(f_A/f_V)^\text{old}$ & $(f_A/f_V)^\text{new}$ & $\mathcal{F}t_0^\text{old}$ & $\mathcal{F}t_0^\text{new}$ \\
    \hline 
    $^{19}$Ne \cite{Combs2020} & 1.0143(29) & 1.0012(2) & 6200(21) & 6142(16)\\
    $^{21}$Na \cite{Vetter2008} & 1.0180(36) & 1.0019(4) & 6179(44) & 6152(42)\\
    $^{29}$P \cite{Masson1990} & 1.0223(45) & 0.9992(1) & 6535(606) & 6496(593)\\
    $^{35}$Ar \cite{Naviliat-Cuncic2009} & 0.9894(21) & 0.9930(14) & 6126(51) & 6135(51)\\
    $^{37}$K \cite{Shidling2014, Fenker2018} & 1.0046(9) & 0.9957(9) & 6141(33) & 6135(33)
    \end{tabular}
    \end{ruledtabular}
    \caption{Difference in calculated $f_A/f_V$ values and its effect on $\mathcal{F}t_0$ for the mirror $T=1/2$ transitions for which all experimental information is available to allow extraction of $|V_{ud}|$. $\mathcal{F}t$ value are taken from \cite{SeverijnsWM} for all isotopes. Uncertainties in $f_A/f_V$ are taken as 20\% of the deviation from unity \cite{Severijns2008}, reflecting an uncertainty in the shell model calculations of a matrix element in $f_A$ \cite{Hayen2018}.}
    \label{tab:new_Vud_mirrors}
\end{table}

It is exactly this Gamow-Teller-specific RC part that is included in the Behrens-B\"uhring part that gives the most significant shift in $f_A/f_V$, which is now removed. The reason why, e.g., the general weak magnetism spectral correction \cite{Hayen2018}, which typically results in a slope of $\sim 0.5\%$ MeV$^{-1}$ for a Gamow-Teller transition, does not contribute can be understood from a theorem by Weinberg \cite{Weinberg1959}. The latter states that - in the absence of QED - no vector-axial vector cross terms can contribute to a scalar quantity such as the lifetime. While the $\gamma W$ box is a dramatic example of when QED does interfere with this theorem, the influence of the weak magnetism spectral correction integrates to zero were it not for the Fermi function. Other spectral features coming from induced currents are seen to have a similar effect in, e.g., the explicit calculation by Wilkinson for the neutron \cite{Wilkinson1982}. The differences between $f_A$ and $f_V$ are now much smaller as finite size corrections are very similar for axial and vector transitions \cite{Hayen2018}. The change in $\mathcal{F}t_0$ is strongest for $^{19}$Ne due to the large value for $\rho$, where the change in $f_A/f_V$ causes a dramatic $3.4 \sigma$ shift in $\mathcal{F}t_0$ and reduces the uncertainty by $24\%$. Given that this is the most accurate determination of $\mathcal{F}t_0$, its influence cannot be understated.

\begin{figure}[!ht]
    \centering
    \includegraphics[width=0.48\textwidth]{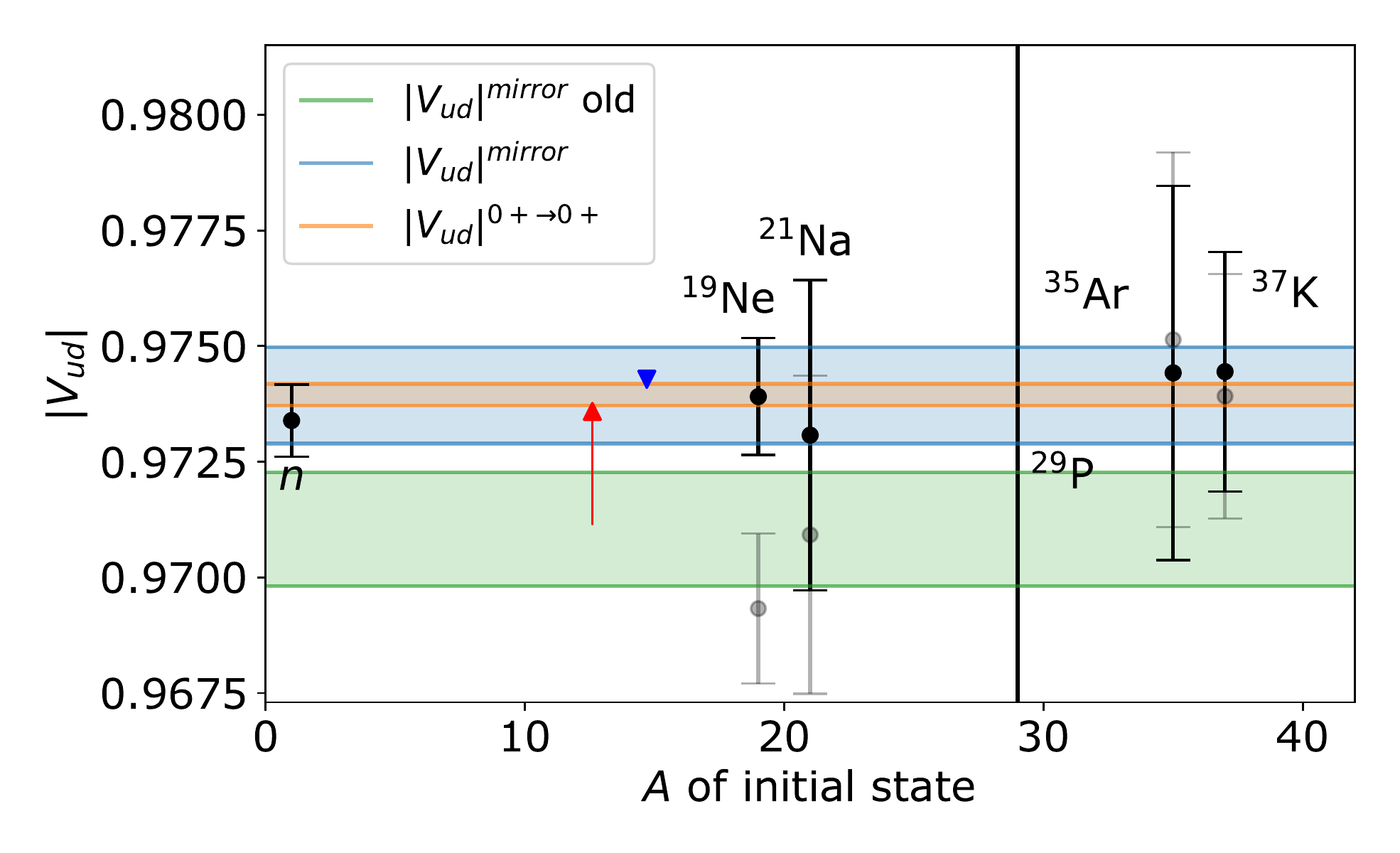}
    \caption{Results with $1\sigma$ uncertainty of $|V_{ud}|$ from mirror decays, superallowed $0^+\to 0^+$ Fermi decays, and the neutron. The shift in the central value of $|V_{ud}|^\mathrm{mirror}$ is shown with a red arrow, while the shift of the new inner RC \cite{Seng2018} is shown with a blue arrow. The results using $(f_A/f_V)^\mathrm{old}$ are shown for each mirror isotope in gray, with their current value in black. The new results solve the long-standing internal discrepancies in the mirror nuclei data set, have a reduced uncertainty and agree extremely well with both superallowed and neutron data.}
    \label{fig:overview_Vud_nuclear}
\end{figure}

Combining all newly calculated results, one obtains an average $\overline{\mathcal{F}t}_0 = 6141(13)$ with $\chi^2/\nu = 0.119$, resulting in an enhanced internal consistency. Application of Eq. (\ref{eq:Vud_mirrors_general}) then leads to a new value for $|V_{ud}|$ extracted from mirror decays
\begin{equation}
    |V_{ud}|^\text{mirror} = 0.9739(10)
\end{equation}
which lies $0.3\%$ ($3\sigma$) higher than the result obtained using the old $f_A/f_V$ values with the most up-to-date experimental input, $|V_{ud}|^\mathrm{mirror}_{old} = 0.9710(12)$, and $0.3\%$ ($2.2\sigma$) higher than the results previously reported in 2009 \cite{Naviliat-Cuncic2009} when accounting for the new radiative corrections \cite{Seng2018}, $|V_{ud}|^\text{mirror}_\text{09} = 0.9712(17)$. Figure \ref{fig:overview_Vud_nuclear} shows an overview of the current status.

Our new result agrees extremely well with that of superallowed Fermi decays, $|V_{ud}|^{0+\to 0+} = 0.97366(16)$ \cite{Hardy2015, Seng2018} and the neutron \cite{Gonzalez-Alonso2018}. Additionally, it resolves the long-standing internal discrepancy in the mirror $\mathcal{F}t_0$ data set, thereby confirming its value and complementarity. As an example, using only the neutron and $^{19}$Ne $\mathcal{F}t_0$ values it is possibly to constrain new tensor interactions in the charged weak current at the $5.1$ TeV level (90\% C.L.) \cite{Combs2020}. Because of the sensitivity enhancement to $\rho$ that several mirror isotopes offer \cite{Hayen2020a}, these present an enticing prospect for complementary study.

\section{Conclusions}
\label{sec:conclusions}
In summary, we presented for the first time a complete calculation of the $\mathcal{O}(\alpha)$ inner radiative corrections to Gamow-Teller transitions. Although \textit{a priori} three contributions specific to the latter compared to Fermi transitions can be identified, two of these depend on the divergence of the axial current and we find that their contribution vanishes in the UV. Additionally, we find that invoking $G$-parity reduces the number of terms in the IR, and their Born contribution vanishes either through crossing symmetry or a cancellation between isoscalar and isovector photon contributions. To $\mathcal{O}(\alpha)$, this leaves the polarized parity-odd contribution of the $\gamma W$ box diagram, analogous to the case of Fermi transitions. We find that the Born contribution is significantly enhanced because of weak magnetism, leading to an increase of a factor 2.9 with respect to Fermi transitions. Following the findings of recent dispersion relation results \cite{Seng2019b}, we take into account additional hadronic contributions besides the Born amplitude below 1 GeV through a model for a holomorphic strong coupling constant. The latter agrees well with experimental data for the polarized Bjorken sum rule, which up to isospin breaking determines the running of the coupling in the vector $\gamma W$ box. For consistency, we treat the axial $\gamma W$ box in the same way, using instead a combination of polarized Bjorken and Gross-Llewellyn Smith sum rules. Even though the latter has limited experimental data available we obtain good agreement using the same model, supplemented by continuity requirements across the threshold. Within uncertainties, this results in the same increase below $\sim 1$ GeV$^2$ as for the vector $\gamma W$ case. In both cases we have discussed higher-twist and target mass corrections, with the latter providing the dominant increase at low $Q^2$. Using these methods, we find $\Delta_R^V = 0.02473(27)$ and $\Delta_R^A = 0.02532(22)$ for a difference $\Delta_R^A - \Delta_R^V = 0.60(4)\times 10^{-3}$. We note that the vector prediction is consistent with both dispersion relation \cite{Seng2019b} and similar recent work \cite{Czarnecki2019}, with the increase with respect to the latter arising from an integration of the Born contribution for $0 \leq Q^2 < \infty$ rather than up to the deep inelastic scattering threshold, and the inclusion of target mass corrections. The difference between vector and axial inner RC is dominated by the weak magnetism Born contribution.

This allowed us to, for the first time, extract the underlying $g_A^{0}$, which is required for use in neutral current processes and compared to lattice QCD. Using the most precise individual measurement, the shift corresponds to $0.7\sigma$. As experimental precision increases further with several upcoming measurements, this effect becomes statistically significant.

We discussed the effect of our findings on an extraction of limits on exotic right-handed currents from comparisons of experimental and lattice QCD $g_A$ determinations. Within the current precision of the latter, the calculated shift is not significant. As the raw $\gamma W$ box integral for the axial vector renormalization is almost $70\%$ larger than the equivalent integral for the vector transition, this difference should be clearly visible when explicitly putting photons on the lattice.

Finally, we explicitly showed how some of the vector-axial vector RC difference is present in some traditional $\beta$ decay formalisms. More importantly, however, we found that some of these contributions cancel in the full $\mathcal{O}(\alpha)$ calculation not present in the traditional results. Additionally, we corrected a double-counting instance in the $|V_{ud}|$ extraction from isospin $T=1/2$ mirror nuclei because of inconsistent experimental extraction and theory input, originating from the partial inclusion of the effect described here. Besides resolving the internal inconsistency in the mirror data set, the extracted $|V_{ud}|^\mathrm{mirror} = 0.9739(10)$ now is excellent agreement with both neutron and superallowed $0^+ \to 0^+$ Fermi determinations. This reinforces the quality of the mirror data set and stresses its potential.

\begin{acknowledgements}
I would like to thank Chien-Yeah Seng for comments and corrections which significantly improved this manuscript. Additionally, I would like to thank Nathal Severijns, Vincenzo Cirigliano, Mikhail Gorchtein, Barry R. Holstein, Albert Young, Andre Walker-Loud and the organizers of ECT*: \textit{Precise beta decay calculations for searches for new physics} and ACFI Amherst: \textit{Current and Future Status of the First-Row CKM Unitarity} workshops for productive discussions related to this manuscript. I would also like to thank Johannes Bl\"umlein for bringing the heavy-flavour corrections to my attention. I acknowledge support by the U.S. National Science Foundation (PHY-1914133), U.S. Department of Energy (DE-FG02-ER41042), the Belgian Federal Science Policy Office (IUAP EP/12-c) and the Fund for Scientific Research Flanders (FWO).
\end{acknowledgements}

\appendix

\section{Interactions with main diagrams}
\label{app:photonic}

\begin{figure}[ht]
    \centering
    \includegraphics[width=0.2\textwidth]{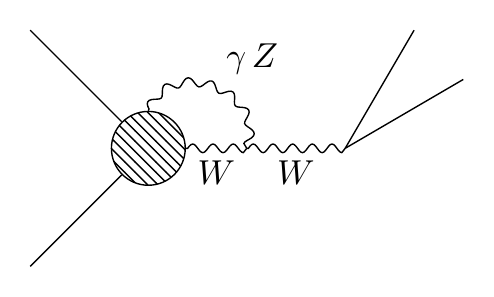}
    \includegraphics[width=0.24\textwidth]{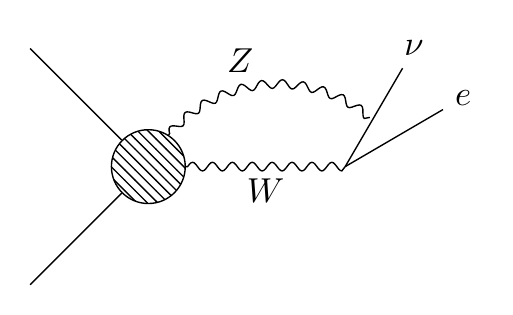}
    \caption{\label{fig:feynman_order_alpha_add_2p}Additional $\mathcal{O}(\alpha)$ two-point correlation function contributions and $ZW$ box diagram that interact with the main diagrams of Fig. \ref{fig:feynman_order_alpha}.}
\end{figure}

While the diagrams shown in Fig. \ref{fig:feynman_order_alpha} are the main contributors for a difference in $\Delta_R^{V, A}$, several of the terms arising from the latter interact with diagrams common to Fermi and Gamow-Teller. To $\mathcal{O}(\alpha)$, the ones important for this work are shown in Fig. \ref{fig:feynman_order_alpha_add_2p}.

Both of these correspond to additional two-current correlation function which are in essence universal to both Fermi and Gamow-Teller transitions, and serve the cancel or combine with elements of the calculation presented above. Taking the virtual $Z$ expanded vertex diagram as an example, the matrix element can be written as
\begin{align}
    \mathcal{M}^Z_1 &= -\frac{ig^4}{2(2\pi)^4}\frac{L^\mu}{q^2-M_W^2}\int \frac{d^4k}{(k^2-M_Z^2)[(k-q)^2-M_W^2]} \nonumber \\
    &\times [(2k-q)_\mu g_{\lambda\rho} + (2q-k)_\lambda g_{\mu\rho} - (k+q)_\rho g_{\mu\lambda}]T^{\lambda\rho}_Z
    \label{eq:M_Z}
\end{align}
The asymptotic behaviour can once again be studied using an OPE or the BJL limit. In the case of the former the leading behaviour for large $k$ is determined by the lowest order operator on the OPE. Because of the charge change in $T^{\lambda\rho}_Z$, this operator must be bilinear in the quark fields. Dimensional analysis learns then that $T^{\lambda\rho}_Z$ behaves as $k^{-1}$ and the integral in Eq. (\ref{eq:M_Z}) is logarithmically divergent for the $k$ products in the numerator, while it is finite but of $\mathcal{O}(G_F^2)$ for the $q$ products because of the presence of the $Z$ mass. Further, we can use the Ward-Takahashi identities for the $k_\lambda$ and $k_\rho$ products. Similar to Eqs. (\ref{eq:WTI_qed}) and (\ref{eq:WTI_weak}), this results in the appearance of Born amplitudes and derivatives in the currents. One can check that the latter contribute only at $\mathcal{O}(G_F^2)$ based on dimensional analysis \cite{Sirlin1978} or brute-force through the BJL limit. One finds then \cite{Sirlin1978}
\begin{align}
    \mathcal{M}^Z_1 &= -\frac{ig^4}{2(2\pi)^4}\frac{L^\mu}{q^2-M_W^2}\int \frac{d^4k}{(k^2-M_Z^2)(k^2-M_W^2)} \nonumber \\
    &\times \biggl[2k_\mu T^{\lambda}_{Z\lambda} + 2i\cos^2 \theta_W \langle p_f | J_\mu^W(0) | p_i \rangle \biggr]
\end{align}
The first term partially cancels the contribution from the vertex correction of Eq. (\ref{eq:derivative_Z}) for $Z$ exchange, and similarly for photon exchange (cfr. Eq. (\ref{eq:deriv_T_PI})). The second term is proportional to the tree-level amplitude and in fact does not depend on the initial and final states. Although the integral is divergent, the latter implies that it is absorbed into the definition of $G_F$ taken from the muon lifetime and we need not worry about it further (see Sec. \ref{sec:outline}). 

The non-asymptotic part of the $ZW$ box diagram contributes only to $\mathcal{O}(G_F^2)$ thanks to the double heavy boson propagator. The asymptotic behaviour of the $ZW$ box diagram is discussed at length in Ref. \cite{Sirlin1978}, and it is - most importantly - to lowest order proportional to the tree-level amplitude and therefore common to Fermi and Gamow-Teller transitions. We merely state the final result
\begin{align}
    \mathcal{M}^Z_2 &= \frac{\alpha}{4\pi}\mathcal{M}_0 \cot^2 \theta_W \left\{2+ \frac{1+R}{1-R}\ln R\right\},
\end{align}
where
\begin{equation}
    R = \frac{M_W^2}{M_Z^2} = \cos^2 \theta_W
\end{equation}
as usual in the Standard Model, and was mentioned in Sec. \ref{sec:outline}.

We have omitted all $\mathcal{O}(\alpha)$ graphs which leave the weak vertex untouched, although their contributions are necessary for the complete calculation. Specifically, the wave function renormalization of the outgoing $\beta$ particle and real bremsstrahlung emission are required for a removal of the infrared divergences appearing in the $\gamma W$ box of Eqs. (\ref{eq:gamma_W_box_def}). Since these are well-known and common to Fermi and Gamow-Teller decays \cite{Sirlin1967}, we do not include a specific discussion.

\section{Born contribution to the $\gamma W$ box}
\label{app:born}
The treatment of the Born contribution proceeds along analogous lines as those described in work by Towner \cite{Towner1992a}. The vertex functions describing the Born couplings of nucleons to electromagnetic and weak fields were given in Eqs. (\ref{eq:EM_nucleon_vertex}) and (\ref{eq:current_decomp_nucleon}). Writing the Born contribution to the $T_{\mu\nu}$ tensor explicitly
\begin{align}
    T_{\mu\nu}^\mathrm{Born} &= \bar{u}(p) \left\{\left[F_1\gamma_\mu +i\frac{F_2}{2M}\sigma_{\mu\lambda}k^\lambda\right]\frac{\slashed{p}-\slashed{k}+M}{k^2-2p\cdot k}\right. \nonumber \\
    &\times\left[g_V\gamma_\nu-i\frac{g_M}{2M}\sigma_{\nu\rho}k^\rho+g_A\gamma_\nu\gamma^5\right]\nonumber \\
    &+\left[g_V\gamma_\nu-i\frac{g_M}{2M}\sigma_{\nu\rho}k^\rho+g_A\gamma_\nu\gamma^5\right]\frac{\slashed{p}+\slashed{k}+M}{k^2+2p\cdot k} \nonumber \\
    &\left.\times \left[F_1\gamma_\mu +i\frac{F_2}{2M}\sigma_{\mu\lambda}k^\lambda\right]\right\}u(p)
    \label{eq:app_born_gammaW}
\end{align}
we take into account only isoscalar photons as discussed above since it is trivial to show that isovector contributions vanish due to crossing symmetry. The renormalization of $g_V$ is affected only by the $g_A$ term, the analysis of which can be found in Ref. \cite{Towner1992a} and more recent work \cite{Seng2019b}. Analogously, $g_A$ is affected only by $g_V$ and - more importantly - $g_M$ and we use
\begin{subequations}
\begin{align}
    \bar{u}(p)&[\gamma_\mu(\slashed{p}-\slashed{k}+M)\gamma_\nu]u(p) \nonumber \\
    &\stackrel{\mathrm{FCC}}{=} -i\epsilon_{\mu\rho\nu\sigma}k^\rho\bar{u}(p)\gamma^\sigma\gamma^5u(p) \\
    \bar{u}(p)&[\sigma_{\mu\alpha}k^\alpha(\slashed{p}-\slashed{k}+M)\gamma_\nu]u(p) \nonumber \\
    &\stackrel{\mathrm{FCC}}{=}0 \\
    \bar{u}(p)&[\gamma_\mu(\slashed{p}-\slashed{k}+M)\sigma_{\nu\alpha}k^\alpha]u(p) \nonumber \\
    &\stackrel{\mathrm{FCC}}{=}2M\epsilon_{\mu\rho\nu\sigma}k^\rho\bar{u}(p)\gamma^\sigma\gamma^5u(p) \\
    \bar{u}(p)&[\sigma_{\mu\beta}k^\beta(\slashed{p}-\slashed{k}+M)\sigma_{\nu\alpha}k^\alpha]u(p) \nonumber \\
    &\stackrel{\mathrm{FCC}}{=}-i(k^2-2k\cdot p)\epsilon_{\mu\rho\nu\sigma}k^\rho \bar{u}(p)\gamma^\sigma\gamma^5 u(p) 
\end{align}
\end{subequations}
where FCC means we only retain terms which transform like first-class currents for an axial transition. The $g_MF_2$ term is suppressed by $1/4M^2$ which we neglect going forward\footnote{Numerically this contribution is less than 1 part in $10^5$.}. If we plug these expressions into Eq. (\ref{eq:app_born_gammaW}) and the integral of Eq. (\ref{eq:gamma_W_box_WTI}), and combine Levi-Civita tensors using $\epsilon_{\mu\rho\nu\sigma}\epsilon^{\mu\lambda\nu\alpha}=-2(\delta^\lambda{}_\rho\delta^\alpha{}_\sigma-\delta^\lambda{}_\sigma\delta^\alpha{}_\rho)$ we find
\begin{align}
    \mathcal{M}_{\gamma W}^\mathrm{Born, A} &=-i2\sqrt{2}\pi \alpha G_FV_{ud}L^\mu \int \frac{d^4k}{(2\pi)^4} \nonumber \\
    &\times [g_VF_1(P_+ + P_-)+g_MF_1P_-+g_VF_2P_+] \nonumber \\
    &\times \bar{u}(p)[k^2\gamma_\mu-\slashed{k}k_\mu]\gamma^5u(p)
    \label{eq:app_born_gamma_W}
\end{align}
with $P_{\pm} = (k^2\pm 2k\cdot p)^{-1}$. The momentum integral is of the form
\begin{equation}
    \int \frac{d^4k}{(2\pi)^4} k_\mu k_\nu F(p\cdot k, k^2) = g_{\mu\nu}I_1 + \frac{p_\mu p_\nu}{M^2}I_2
    \label{eq:app_integral_decomp}
\end{equation}
due to Lorentz covariance for a general scalar function $F$. Plugging this into Eq. (\ref{eq:app_born_gamma_W}) results in
\begin{align}
    \mathcal{M}_{\gamma W}^\mathrm{Born, A} &=-i2\sqrt{2}\pi \alpha G_FV_{ud} \nonumber \\
    &\times\bar{u}(p)\left[3I_1\gamma_\mu+I_2\left(\gamma_\mu+\frac{p_\mu}{M}\right)\right]\gamma^5u(p)L^\mu
\end{align}
which is similar in form to what is found in Ref. \cite{Towner1992a} for the vector case. In the latter, the main correction stems from the timelike contribution for which the $I_2$ prefactor is $\mathcal{O}(q^2/M^2)$. For the spacelike contribution to axial vector transition we have $p_i/M = \mathcal{O}(q/M)$ and $\bar{u}\gamma^5u = \mathcal{O}(q/M)$ so that the $I_2$ integral contributes to leading order. The two integrals can be found easily from Eq. (\ref{eq:app_integral_decomp})
\begin{align}
    I_1 &= \frac{1}{3}\int \frac{d^4k}{(2\pi)^4}(k^2-\nu^2)F(p\cdot k, k^2) \\
    I_2 &=\frac{1}{3}\int \frac{d^4k}{(2\pi)^4}(4\nu^2-k^2)F(p\cdot k, k^2).
\end{align}
The integrals can be brought into the $Q^2=-k^2$ variable through a Wick rotation and using
\begin{align}
    \int \frac{d^4k}{(2\pi)^4}F(\nu, Q^2) &= \frac{i}{8\pi^3}\int_0^\infty dQ^2 Q^2 \nonumber \\
    &\times\int_{-1}^1du \sqrt{1-u^2}F(iQu, Q^2)
\end{align}
with $\nu=p\cdot k/M$ as before. Putting everything together, we find
\begin{align}
    &\mathcal{M}_{\gamma W}^\mathrm{Born, A} = -\frac{\sqrt{2}\alpha}{4\pi}G_FV_{ud}\bar{u}(p)\gamma_\mu\gamma^5u(p)L^\mu\nonumber \\
    &\times \int \frac{dQ^2}{Q^2} \frac{5+4r}{3(1+r)^2}\left[g_V\left(F_1+\frac{F_2}{2}\right)+\frac{g_MF_1}{2}\right]
\end{align}
with
\begin{equation}
    r = \sqrt{1+4M^2/Q^2}.
\end{equation}
Comparing to the leading order expression for the axial transition we can write
\begin{align}
    \Box_{VV}^\mathrm{Born} &= -\frac{\alpha}{2\pi} \int \frac{dQ^2}{Q^2}\frac{5+4r}{3(1+r)^2} \nonumber \\
    &\times\frac{g_V\left(F_1+\dfrac{F_2}{2}\right)+\dfrac{g_MF_1}{2}}{g_A(0)}
\end{align}
leading to the expressions in the main text (keeping in mind $g_A(0) < 0$ in our definition).

We perform the integration by defining the form factors as $g_i(Q^2) = g_i(0)G_i(Q^2)$. If we assume a standard dipole form ${G_D(Q^2) = (1-Q^2/\Lambda^2)^{-2}}$, these expressions can be put into closed form using standard methods \cite{Fukugita2004}. Instead, we follow Ref. \cite{Seng2019b} and use the global fit results of Ref. \cite{Ye2018} for the Sachs isoscalar magnetic moment and the vector form factor, and Ref. \cite{Bhattacharya2011} for the axial form factor. Invoking the conserved vector current hypothesis, we use the isovector magnetic moment also for $G_M(Q^2)$. The numerical results are summarized in Eqs. (\ref{eq:box_F_Born})-(\ref{eq:box_GT_Born}).

\section{Deep inelastic scattering and QCD sum rules}

The deep inelastic scattering (DIS) contribution to the $\gamma W$ box diagram was discussed in terms of different QCD sum rules. Here, we summarize the main results.

\subsection{Axial vector transition}
Like the famous axial vector contribution to the $\gamma W$ box for Fermi transitions, an analogous situation occurs for the axial transition with the isoscalar photonic and weak vector current, shown in Eq. (\ref{eq:V_munu}). The OPE expression discussed in the main text is proportional to the tree-level amplitude, but contains perturbative QCD (pQCD) corrections. We can relate these corrections quite easily to those of the polarized Bjorken sum rule, which treats the Cornwall-Norton moments of the polarized $g_1$ function in proton and neutron, i.e.
\begin{equation}
    \int_0^1 dx [g_1^p(x)-g_1^n(x)] = \frac{1}{6}\left|\frac{g_A}{g_V} \right|\left(1-\frac{\alpha_{g_1}(Q^2)}{\pi}\right)
\end{equation}
where the constant prefactor can be determined using current algebra or the quark parton model. The pQCD corrections can be determined by using the operator product expansion of
\begin{align}
    &i\int dz \exp(iqz) T\{V_\mu^a(z) V_\nu^b(0)\} \nonumber \\ 
    \stackrel{Q^2\to \infty}{\simeq} &\epsilon_{\mu\nu\rho\sigma} \frac{q^\sigma}{q^2}C^{(A)}(\mu^2/Q^2, \alpha_s) d^{abc}A^\rho_c(0) + \ldots
    \label{eq:app_BjSR_OPE}
\end{align}
where $V_\mu^a = \overline{\psi} \gamma_\mu t^a \psi$ and $A^a_\mu = \overline{\psi} \gamma_\mu \gamma_5 t^a \psi$ are non-singlet vector and axial-vector quark currents, respectively, with $t^5$ the $SU(3)$ flavour generators \cite{Larin1991}. For simplicity, we take only $u, d, s$ quarks into account. In order to relate it to our $\gamma W$ diagram, we define the relevant currents
\begin{subequations}
 \begin{align}
    V^\pm &= \frac{1}{\sqrt{2}}(V^1\pm iV^2) \\
    A^\pm &= \frac{1}{\sqrt{2}}(A^1\pm iA^2) \\
    J^\gamma &= V_3 + \frac{1}{\sqrt{3}}V_8
\end{align}
\end{subequations}
is the $SU(3)$ flavour representation \cite{Adler1968, Treiman1972}. For the isoscalar photonic contribution only $V_8$ contributes, and using $d^{811} = d^{822} = 1/\sqrt{3}$
\begin{align}
&i\int dz \exp(iqz) T\{V_\mu^\pm(z) J^\gamma_{S}(0)\} \nonumber \\
= &\frac{i}{\sqrt{3}}\int dz \exp(iqz) T\{V_\mu^\pm(z) V_\nu^8(0)\} \nonumber \\
\stackrel{Q^2\to \infty}{\simeq} &\epsilon_{\mu\nu\rho\sigma} \frac{q^\sigma}{q^2\sqrt{3}}C^{(A)}(\mu^2/Q^2, \alpha_s) A^\rho_\pm(0)
\end{align}
and so regardless of $SU(3)$ breaking the pQCD corrections to the $\gamma W$ box are exactly those of the polarized Bjorken sum rule, if we neglect contributions from strange quarks present in $V_8$.

\subsection{Vector transition}
In the case of the $\gamma W$ box contribution to the vector transition, the situation is somewhat more complex. The original idea by Marciano and Sirlin \cite{Marciano2006, SengPC} was to relate the axial vector contribution to the Bjorken sum rule through a chiral rotation, i.e. $d \to \gamma_5 d$ and $s \to \gamma_5 s$. Above $\Lambda_{\chi}$ the Standard Model Lagrangian is invariant under such chiral transformations, and the electromagnetic current is unchanged while transforming $V^\pm_\mu \to A^\pm_\mu$ and vice versa. Reference \cite{Czarnecki2019} took this approach one step further and used the polarized Bjorken sum rule data also at low momenta (i.e. $Q^2 \ll \Lambda_{\chi}^2$) to describe the low and intermediate momenta contributions not captured by the elastic channel. As the approximation of chiral invariance breaks down below this scale, however, the correspondence is not rigorously expected to hold. In their work, this low-$Q^2$ region contributes $4.6(9) \times 10^{-4}$ to $\Delta_R^V$, with a generic $20\%$ uncertainty. It is currently not clear whether this corresponds to an over- or underestimation of the true uncertainty. 

Another approach that was discussed briefly in Ref. \cite{Seng2019b} was to, besides explicit modeling, relate the behaviour of the axial $\gamma W$ contribution to charged current (anti)neutrino-nucleon scattering. The structure functions probed in the latter obey the GLS sum rule,
\begin{equation}
    \int_0^1 dx[F_3^{\bar{\nu}}+F_3^{\nu}] = 6\left[1-\frac{\alpha_{F_3}(Q^2)}{\pi}\right]
\end{equation}
with $F_3$ the parity-violating structure function similar to the ones discussed in the main text. The pQCD corrections can be similarly obtained from an OPE \cite{Larin1991}
\begin{align}
    &i\int d^4z \exp(iqz) T\{A^a_\mu(z)V^b_\mu \} \nonumber \\
    \stackrel{Q^2\to \infty}{\simeq} &\delta^{ab}\epsilon_{\mu\nu\rho\lambda} \frac{q^\lambda}{q^2}C^{(V)}(\mu^2/Q^2, \alpha_s)V^\kappa(0) + \ldots
\end{align}
with definitions equivalent to Eq. (\ref{eq:app_BjSR_OPE}), and $V^\kappa = \overline{\psi}\gamma^\kappa \psi$ is a singlet vector current. The Kronecker delta makes identification with the axial vector $\gamma W$ box contribution less obvious. In fact, connection with the isoscalar electromagnetic contribution ($V^8_\mu$) is impossible in this form without resorting to a chiral transformation as above. Instead, we continue with the \textit{isovector} part of the electromagnetic current, $V^3_\mu$, and relate it to $A^3_\mu$
\begin{align}
    &i\int d^4z \exp(iqz) T\{A^3_\mu(z)J^\gamma_{V}{}_{\nu}(0) \} \nonumber \\
    \stackrel{Q^2\to \infty}{\simeq} &\epsilon_{\mu\nu\rho\lambda} \frac{q^\lambda}{q^2}C^{(V)}(\mu^2/Q^2, \alpha_s)V^\kappa(0) + \ldots
\end{align}
If we choose a representation of $SU(3)$ such that
\begin{equation}
    t^i = \frac{1}{2}\left(\begin{array}{cc}
        \tau^i & 0 \\
        0 & 0
    \end{array} \right)
\end{equation}
where $i=1,2,3$ and $\tau^i$ are the $SU(2)$ Pauli matrices, we can relate $A^3_\mu$ to $A^\pm_\mu$ using isospin symmetry. If we then assume isoscalar and isovector behaviour is sufficiently similar, we can up to additional isospin symmetry breaking corrections use the GLS pQCD corrections for those of the axial $\gamma W$ box contribution. Reference \cite{Seng2019b} found a similar isovector-isoscalar correspondence in the Born channel, and argued that the $I=0$ and $I=1$ Regge physics at intermediate scales can be easily related, albeit with the model-dependence inherent to the Regge description. Since isospin is broken at a lower scale than chiral symmetry, however, we believe the use of the GLS Nachtmann moment can be more easily defended at low $Q^2$.

\bibliography{library}
\end{document}